\documentclass[english,runningheads,11pt]{llncs}
\usepackage{tabularx,booktabs,multirow,delarray,array}
\usepackage{graphicx,amssymb,amsmath,amssymb}
\usepackage{latexsym}

\usepackage[in]{fullpage}

\newcommand{\bcls}{\mbox{MSBC}}
\newcommand{\msbc}{\mbox{MSBC}}

\DeclareMathOperator*{\argmin}{arg\,min}

\newenvironment{myproof}{\noindent {\textbf{Proof:}}\rm}{\hfill $\Box$\rm}

\begin{document}

\title{Minimizing the Aggregate Movements for Interval Coverage
}

\author{Aaron M. Andrews
\and
Haitao Wang
}

\institute{
Department of Computer Science\\
Utah State University, Logan, UT 84322, USA\\
\email{aaron.andrews@aggiemail.usu.edu, haitao.wang@usu.edu}\\
}

\maketitle

\pagestyle{plain}
\pagenumbering{arabic}
\setcounter{page}{1}

\begin{abstract}
We consider an interval coverage problem.
Given $n$ intervals of the same length on a line $L$ and a line segment $B$
on $L$, we want to move the intervals along $L$ such
that every point of $B$ is covered by at least one interval and the sum
of the moving distances of all intervals is minimized.
As a basic geometry problem, it has
applications in mobile sensor barrier coverage in wireless sensor
networks. The previous work solved the problem in $O(n^2)$ time.
In this paper, by discovering many interesting observations and
developing new algorithmic techniques, we present an $O(n\log
n)$ time algorithm. We also show an $\Omega(n\log n)$ time lower bound
for this problem, which implies the optimality of our algorithm.
\end{abstract}

\section{Introduction}
\label{sec:intro}

In this paper, we consider an interval coverage problem.
Given $n$ intervals of the same length on a line $L$ and a line segment $B$
on $L$, we want to move the intervals along $L$ such
that every point of $B$ is covered by at least one interval and the sum
of the moving distances of all intervals is minimized.

The problem has applications in barrier coverage of mobile sensors in wireless sensor
networks. For convenience, we will introduce and discuss the problem
from the barrier coverage point of view.
Given a set of $n$ points $S=\{s_1,s_2,\ldots,s_n\}$ on
$L$, say, the $x$-axis, each point $s_i$ represents a
sensor. Let $x_i$ be the coordinate of $s_i$ on $L$ for each $1\leq
i\leq n$. For any two coordinates $x$ and $x'$ with $x\leq x'$, we use
$[x,x']$ to denote the interval of $L$ between $x$ and $x'$.
The sensors of $S$ have the same {\em covering range}, denoted by $z$, such that
for each $1\leq i\leq n$, sensor $s_i$ {\em covers} the interval
$[x_i-z,x_i+z]$. Let $B$ be a line segment of $L$ and we call $B$ a
``barrier''. We assume that the length of $B$ is no more than $2z\cdot
n$ since otherwise $B$ could not be fully covered by these sensors. The problem is to move all sensors along $L$ such that each point
of $B$ is covered by at least one sensor of $S$ and the sum of the
moving distances of all sensors is minimized. Note that although
sensors are initially on $L$, they may
not be on $B$. We call this problem the {\em
min-sum barrier coverage}, denoted by \msbc.

The problem \msbc\ has been studied before and Czyzowicz {\em et al.}
\cite{ref:CzyzowiczOn09} gave an $O(n^2)$ time algorithm. In this
paper, we present an $O(n\log n)$ time algorithm and we show that
our algorithm is optimal.

\subsection{Related Work}

A Wireless Sensor Network (WSN) uses a large number of sensors to
monitor some surrounding environmental phenomena
\cite{ref:AkyildizWi02}.
Each sensor is equipped with a sensing device with limited battery-supplied energy.
The sensors process data obtained and forward the data to a base station.
Intrusion detection and border surveillance constitute a
major application category for WSNs. A
main goal of these applications is to detect intruders as they
cross the boundary of a region or domain. For example, research efforts were made to
extend the scalability of WSNs to the monitoring of international
borders \cite{ref:Hu08,ref:KumarBa07}.
Unlike the traditional {\it full coverage}
\cite{ref:LiLo08,ref:YangSc07,ref:ZouA05} which requires an entire target
region to be covered by the sensors,
the {\it barrier coverage}
\cite{ref:BhattacharyaOp09,ref:ChenDe07,ref:CzyzowiczOn09,ref:CzyzowiczOn10,ref:KumarBa07}
only seeks to cover the perimeter of the region to ensure that any
intruders are detected as they cross the region border.
Since barrier coverage requires fewer sensors, it is often preferable to
full coverage. Because sensors have limited
battery-supplied energy, it is desired to minimize their movements.

If the sensors have different ranges, the Czyzowicz {\em et al.}
\cite{ref:CzyzowiczOn10} proves that the problem \msbc\ is NP-hard.

The {\em min-max} version of \msbc\ has also been studied, where the objective is to minimize the maximum movement of all sensors. If the sensors have the same range,
Czyzowicz {\em et al.} \cite{ref:CzyzowiczOn09} gave an $O(n^2)$ time algorithm, and
later Chen {\em et al.} presented an $O(n\log n)$ time solution
\cite{ref:ChenAl13}. If sensors have different ranges, Czyzowicz {\em et al.} \cite{ref:CzyzowiczOn09} left it as an open question whether the problem is NP-hard, and Chen
{\em et al.} \cite{ref:ChenAl13} answered the open problem by giving an
 $O(n^2\log n)$ time algorithm.

Mehrandish {\em et al.}~\cite{ref:MehrandishOn11,ref:MehrandishMi11}
considered another variant of the one-dimensional barrier coverage problem, where the goal is to move the minimum number of
sensors to form a barrier coverage. They  \cite{ref:MehrandishOn11,ref:MehrandishMi11} proved
the problem is NP-hard if sensors have different ranges and gave polynomial time
algorithms otherwise.
In addition, Li {\em et al.}~\cite{ref:LiMi11} considers the
linear coverage problem which
aims to set an energy for each sensor to form a coverage such that the
cost of all sensors is minimized. There \cite{ref:LiMi11}, the sensors
are not allowed to move, and the more energy a sensor has, the larger
the covering range of the sensor and the larger the cost of the
sensor. Another problem variation is considered in \cite{ref:Bar-NoyMa13}, where the goal is to maximize the barrier coverage lifetime subject to the limited battery powers.

Bhattacharya {\em et al.}~\cite{ref:BhattacharyaOp09} studied a
two-dimensional barrier coverage in which the barrier is a
circle and the sensors, initially located inside the circle, are
moved to the circle to minimize the sensor movements; the ranges of the
sensors are not explicitly specified but the destinations of the sensors are
required to form a regular $n$-gon on the circle. Algorithms for both min-sum and min-max versions were given in \cite{ref:BhattacharyaOp09} and subsequent improvements were made in
\cite{ref:ChenOP11,ref:TanNe10}.

Some other barrier coverage problems have been studied.
For example, Kumar {\em et al.}~\cite{ref:KumarBa07} proposed
algorithms for determining whether a region is barrier covered after
the sensors are deployed. They considered both the deterministic
version (the sensors are deployed deterministically) and the
randomized version (the sensors are deployed randomly), and aimed
to determine a barrier coverage with high probability.
Chen {\it et al.}~\cite{ref:ChenDe07} introduced a local barrier coverage
problem in which individual sensors determine the barrier coverage locally.

\subsection{Our Approaches}

If the covering intervals of all sensors intersect the barrier $B$, we call this case the {\em containing case}. If the sensors whose covering intervals do not intersect $B$ are all in one side of $B$, then it is called the {\em one-sided case}. Otherwise, it is the {\em general case}.

In Section \ref{sec:pre}, we introduce notations and briefly review the algorithm in \cite{ref:CzyzowiczOn09}. Based on the algorithm in \cite{ref:CzyzowiczOn09}, by using a different implementation and designing efficient data structures, we give an $O(n\log n)$ time algorithm for the containing case in Section \ref{sec:contain}.

To solve the one-sided case, the containing case algorithm does not work and we have to develop different algorithms. To do so, we discover a number of interesting observations on the structure of the optimal solution, which allows us to have an $O(n\log n)$ time algorithm. The one-sided case algorithm uses the containing case algorithm as a first step and apply a sequence of so-called ``reverse operations''. The one-sided case is discussed in Section \ref{sec:onesided}.

In Section \ref{sec:general}, we solve the general case in $O(n\log n)$ time. To this end, we generalize the techniques for solving the one-sided case. For example, we show a monotonicity property of one-sided case (in Section \ref{sec:onesided}), which is quite useful for the general case. We also discover new observations on the solution structures. These observations help us develop efficient algorithmic techniques. All these efforts lead to the $O(n\log n)$ time algorithm for the general case.

Section \ref{sec:conclude} concludes the paper, where we prove the $\Omega(n\log n)$ time lower bound (even for the containing case) by an easy reduction from sorting.

We should point out that although the paper is relatively long, the algorithm itself is simple and easy to implement. In fact, the most complicated data structure used in the algorithm is the balanced binary search trees! The lengthy (and sometimes tedious) proofs are all devoted to discovering the observations and showing the correctness, which eventually lead to a simple, elegant, efficient, and optimal algorithm. Discovering these observations turns out to be quite challenging and is actually one of our main contributions.


\section{Preliminaries}
\label{sec:pre}

In this section, we introduce some notations and sketch the algorithm given by Czyzowicz {\em et al.}
\cite{ref:CzyzowiczOn10}.
Below we will use the terms ``line segment''
and ``interval'' interchangeably, i.e., a line segment of $L$ is also
an interval and vice versa. Let $\beta$ denote the length of $B$.
Without loss of generality, we assume the
barrier $B$ is the interval $[0,\beta]$.
For short, sensor covering intervals are called {\em sc-intervals}.

We assume the sensors of $S$ are already sorted,
i.e., $x_1\leq x_2\leq \cdots\leq x_n$ (otherwise we sort them in $O(n\log
n)$ time). For each sensor $s_i$, we use
$I(s_i)$ to denote its covering interval. Recall that $z$ is the covering range of each sensor and the length of each sc-interval is $2z$. We assume $2z< \beta$ since otherwise
the solution would be trivial.
An easy but important observation given in \cite{ref:CzyzowiczOn10} is the
following {\em order preserving property:} there always exists an
optimal solution where the order of the sensors is the same as that in
the input. Note that this property does not hold if sensors have
different ranges.


Sensors will be moved during the algorithm. For any sensor
$s_i$, suppose its location at some moment is $y_i$; the value
$x_i-y_i$ is called the {\em displacement} of $s_i$ (here we use $x_i-y_i$ instead of $y_i-x_i$ in the definition in order to ease the discussions later). Hence, if the
displacement of $s_i$ is positive (resp., negative),
then it is to the left (resp., right) of its original location in the input.

In the sequel, we define two important concepts: {\em gaps} and {\em overlaps}, which were also used in \cite{ref:CzyzowiczOn10}.

\begin{figure}[t]
\begin{minipage}[t]{\linewidth}
\begin{center}
\includegraphics[totalheight=0.8in]{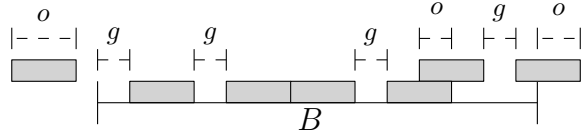}
\caption{\footnotesize Illustrating gaps (denoted by $g$) and overlaps (denoted by $o$).}
\label{fig:Fig_4}
\end{center}
\end{minipage}
\vspace*{-0.15in}
\end{figure}

A \emph{gap} refers to a maximal sub-segment of $B$ such that each
point of the sub-segment is not covered by any sensors (e.g., see
Fig.~\ref{fig:Fig_4}). Each endpoint of any gap is an endpoint of either an sc-interval or $B$.
Specifically, consider two adjacent sensors $s_i$ and $s_{i+1}$ such
that $x_i+z < x_{i+1}-z$. If $0\leq x_i+z$ and $x_{i+1}-z \leq
\beta$, then the interval $[x_i+z,x_{i+1}-z]$ is on $B$ and defines a gap, and
$s_i$ and $s_{i+1}$ are called the left
and right {\em generators} of the gap, respectively.
If $x_i+z < 0 < x_{i+1}-z\leq \beta$, then $[0,x_{i+1}-z]$ is a gap
and $s_{i+1}$ is the only {\em generator} of the gap.
Similarly, if $0\leq x_i+z < \beta < x_{i+1}-z$, then $[x_{i}+z, \beta]$ is a gap
and $s_{i}$ is the only {\em generator}.
For any gap $g$, we use $|g|$ to denote its length.
For simplicity, if a gap $g$ has only one generator $s_i$, then
the left/right generator of $g$ is $s_i$.

\begin{figure}[t]
\begin{minipage}[t]{\linewidth}
\begin{center}
\includegraphics[totalheight=0.8in]{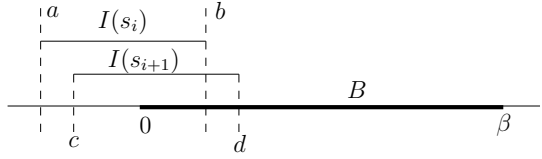}
\caption{\footnotesize $I(s_i)\cap I(s_{i+1})$ contains $0$ in its interior. In this case, we consider $s_i$ and $s_{i+1}$ together defining an overlap $[c,b]$ and $s_{i}$ itself defining an overlap $[a,0]$.}
\label{fig:overlap}
\end{center}
\end{minipage}
\vspace*{-0.15in}
\end{figure}

To solve the problem \bcls, the essential task is to move the sensors
to cover all gaps by eliminating {\em overlaps}, defined as follows.
Consider two adjacent sensors $s_i$ and $s_{i+1}$.
The intersection $I(s_i)\cap I(s_{i+1})\cap B$ defines an overlap if it is not empty (e.g.,
see Fig.~\ref{fig:Fig_4}), and we call $s_i$ and $s_{i+1}$ the left
and right {\em generators} of the overlap, respectively. Consider
any sensor $s_i$. If $I(s_i)$ is not completely on $B$, then the
sub-interval of $I(s_i)$ that is not on $B$ defines an
overlap and $s_i$ is its only generator (e.g., see Fig.~\ref{fig:Fig_4}).
A subtle situation appears when $I(s_i)\cap I(s_{i+1})$ contains
an endpoint of $B$ in its interior. Refer to Fig.~\ref{fig:overlap} as
an example, where $0$ is in the interior of $I(s_i)\cap I(s_{i+1})$ with
$I(s_i)=[a,b]$ and $I(s_{i+1})=[c,d]$. According to our definition, $s_i$ and $s_{i+1}$ together define an overlap $[0,b]$; $s_i$ itself defines an overlap
$[a,0]$; $s_{i+1}$ itself defines an overlap $[c,0]$. However, to avoid some tedious discussions, we consider the union of $[c,0]$ and $[0,b]$ as a single overlap $[c,b]$ defined by $s_i$ and $s_{i+1}$ together, but $s_{i}$ still itself defines the overlap $[a,0]$. Symmetrically, if $I(s_i)\cap I(s_{i+1})$ contains $\beta$ in its interior, then we consider $I(s_i)\cap I(s_{i+1})$ as a single overlap defined by $s_i$ and $s_{i+1}$, and $s_{i+1}$ itself defines an overlap that is the portion of $I(s_{i+1})$ outside $B$.

For any overlap $o$, we use $|o|$ to denote its length.
For simplicity, if an overlap $o$ has only one generator $s_i$, then
the left/right generator of $o$ is $s_i$.
We should point out that according to our above definition on overlaps, if an overlap has two different generators, then these two generators must be two adjacent sensors (e.g., $s_i$ and $s_{i+1}$ for some $i$). In other words, if the sc-intervals of two non-adjacent sensors (e.g., $s_i$ and $s_{i+2}$) intersect, their intersection does not define any overlap.

Clearly, the total number of overlaps and gaps is $O(n)$.

To solve \bcls, the
goal is to move the sensors to cover all gaps by eliminating overlaps.
We say a gap/overlap $go_1$ is to the {\em left} (resp., {\em right}) of another gap/overlap $go_2$
if the left generator of $go_1$ is to the left (resp., right) of the left generator of $go_2$ (in the case of Fig.~\ref{fig:overlap}, where overlaps $[c,b]$ and $[a,0]$ have the same left generator $s_i$, $[a,0]$ is considered to the left of $[c,b]$).


For any two indices $i$ and $j$ with $i\leq j$, let $S(i,j)=\{s_i,
	s_{i+1}, \ldots, s_j\}$.

Below we sketch the $O(n^2)$ time algorithm in
\cite{ref:CzyzowiczOn10} on the containing case where every sc-interval
intersects $B$. The algorithm ``greedily''
covers all gaps from left to right one by one. Suppose the
first $i-1$ gaps have just been covered completely and the algorithm
is about to cover the gap $g_i$.

Let $o_{i}^r$ (resp., $o_i^l$) be
the closest overlap to the right (resp., left) of $g_i$.
We will cover $g_i$ by using either $o_i^r$ or $o_i^l$.
To determine using which overlap to cover $g_i$, the costs
$C(o_i^r)$ and $C(o_i^l)$ are defined as follows.
Let $S_r(g_i)$ be the set of sensors between the right generator of $g_i$
and the left generator of $o_i^r$. Define
$C(o_i^r)$ to be $|S_r(g_i)|$. The intuition of this definition is that
suppose we shift all sensors of $S_r(g_i)$ to the left
for an infinitesimal distance $\epsilon>0$ (such that the gap $g_i$
becomes $\epsilon$ shorter), then the sum of the moving distances of all
sensors of $S_r(g_i)$ is $\epsilon\cdot C(o_i^r)$.
As will be clear later, the current
displacement of each sensor in $S_r(g_i)$ may be positive but cannot be negative.
For $C(o_i^l)$, it is defined in
a slightly different way. Let $S_l(g_i)$ be the set of sensors between the
left generator of $g_i$ and the right generator of $o_i^l$, and let
$S_l'(g_i)$ be the subset of sensors of $S_l(g_i)$ whose
displacements are positive. If we shift all sensors in $S_l(g_i)$ to the right
for an infinitesimal distance $\epsilon>0$, although the sum of
the moving distances of all sensors of $S_l(g_i)$
is $\epsilon\cdot |S_l(g_i)|$, the total moving distance contributed
to the sum of the moving distances of all sensors of $S$ is actually
$\epsilon\cdot (|S_l(g_i)|-2\cdot |S'_l(g_i)|)$ because the sensors of
$S'_l(g_i)$ are moved towards their original locations. Hence, the
cost $C(o_i^l)$ is defined to be $|S_l(g_i)|-2\cdot |S'_l(g_i)|$.
Note that the sensors in $S_r(g_i)$ or $S_l(g_i)$ are consecutive in their index order.

If $C(o_i^r)< C(o_i^l)$, we move each sensor in $S_r(g_i)$ leftwards by
distance $\min\{|o_i^r|,|g_i|\}$, and we call this a {\em left-shift process}. Note that if there is any gap $g_j$ between two sensors in $S_r(g_i)$, then the above shift process will move $g_j$ leftwards as well, but the size and the generators of $g_j$ do not change, and thus in the later algorithm we can still use $g_j$ without causing any problems.  If $|g_i|\leq |o_i^r|$, then
after the left-shift process $g_i$ is covered
completely and we proceed on the next gap $g_{i+1}$.
Otherwise, $o_i^r$ is eliminated and $g_i$ is only partially covered. We proceed on the remaining $g_i$.

If $C(o_i^r)\geq C(o_i^l)$, we move each sensor in $S_l(g_i)$ rightwards by
distance $\min\{|o_i^l|,|g_i|,\alpha\}$, where $\alpha$ is the
smallest displacement of the sensors in $S_l'(g_i)$, and we call this a {\em right-shift process}.
If $\alpha$ is the smallest among the three values, then the process makes the
displacement of at least one sensor in $S'_l(g_i)$ become zero and we
call the process as a {\em positive-displacement-removal} right-shift process (or
PDR process for short). After the process, if $g_i$ is only partially covered,
we proceed on the remaining $g_i$; otherwise
we proceed on the next gap $g_{i+1}$.

The algorithm finishes after all gaps are covered. To analyze the running time, there are $O(n)$ shift processes in total.
To see this, each shift process covers a gap completely, or eliminates an overlap,
or is a PDR process. An observation is that if the displacement of a
sensor $s_i$ was positive but is made to zero during a PDR process, then the
displacement of $s_i$ will never become positive again because
all uncovered gaps are
to the right of $s_i$. Therefore, the number of PDR processes is at most
$n$. Since the number of gaps and overlaps is $O(n)$, the total number
of shift processes in the algorithm is $O(n)$. Each shift process can be done in
$O(n)$ time, and thus the algorithm runs in $O(n^2)$ time.

\section{The Containing Case}
\label{sec:contain}

In this section, we present our algorithm that solves the containing case of \bcls\ in $O(n\log n)$ time. The high-level scheme of our algorithm is the same as the $O(n^2)$ time algorithm \cite{ref:CzyzowiczOn10} described in Section \ref{sec:pre}, but we design efficient data structures such that each shift process can be implemented in $O(\log n)$ amortized time. More specifically, our algorithm maintains an {\em overlap tree} $T_o$, a {\em position tree} $T_p$, a {\em left-shift tree} $T_l$, and a global variable $\gamma$.

\subsection{The Overlap Tree $T_o$}

We store each gap/overlap by recording its generators.
Consider any gap $g_i$ (which may have been partially covered previously). Our
algorithm needs to compute the two overlaps $o_i^l$ and $o_i^r$.
To this end, we maintain all overlaps in a balanced
binary search tree $T_o$, called {\em overlap tree}, using the indices of the left generators of the overlaps as ``keys''.
We can find the two overlaps $o_i^l$ and $o_i^r$ in $O(\log n)$ time by searching $T_o$ with
the index of the left generator of $g_i$. The tree $T_o$ can also support each deletion of any overlap in $O(\log n)$ time if the overlap is eliminated.

Furthermore, $T_o$ can
help us to compute the costs $C(o_i^l)$ and $C(o_i^r)$ in the
following way. After $o_i^r$ is found, we have $|S_i^r|=a-b+1$, where
$a$ is the index of the left generator of
$o_i^r$ and $b$ is the index of the right generator of $g_i$.
Hence, $C(o_i^r) = |S_r(g_i)|$ can be computed in
$O(\log n)$ time. Similarly, we can obtain $|S_l(g_i)|$. However, to
compute $C(o_i^l)$, we also need to know the size $|S'_l(g_i)|$, which
will be discussed later.

\subsection{The Position Tree $T_p$}

Recall that the algorithm needs to do the left or right shift processes,
each of which moves a sequence of consecutive sensors by the same distance. To
achieve the overall $O(n\log n)$ time for the algorithm, we
cannot explicitly move the involved sensors for each shift process. Instead, we
use the following {\em position tree} $T_p$ to perform each shift
implicitly in $O(\log n)$ time.

The tree $T_p$ is a complete binary tree of $n$
leaves and $O(\log n)$ height. The leaves from left to right correspond to the sensors in their index order. For each $1\leq j\leq n$, leaf $j$ (i.e., the $j$-th leaf from the left) stores the original location $x_j$ of sensor $s_j$. Each node of $T_p$ (either an internal node or a leaf) is associated with a {\em shift} value. Initially the shift values of all
nodes of $T_p$ are zero. At any moment during the algorithm, the actual
location of each sensor $s_j$ is $x_j$ plus the sum of the shift
values of the nodes in the path from the root to leaf $j$ (actually this sum of shift values is exactly the negative value of the current displacement of $s_j$), which
can be obtained in $O(\log n)$ time.

Now suppose we want to do a right-shift process that moves a sequence of
sensors in $S(j,k)$ for $j\leq k$ rightwards by a distance $\delta$.
We first find a set $V_{jk}$ of $O(\log n)$ nodes of $T_p$ such that the
leaves of the subtrees of all these nodes correspond to exactly the sensors in
$S(j,k)$. Specifically, $V_{jk}$ is defined as follows. Let $w$ be the lowest common ancestor of leaves $i$ and $j$. Let $\pi'_j$ be the path from the parent of leaf $j$ to $w$. For each node $v$ in
$\pi'_j$, if the right child of $v$ is not in $\pi'_j$, then the right child of $v$ is in $V_{jk}$.
Leaf $j$ is also in $V_{jk}$. The rest of the nodes of $V_{jk}$ are defined in a symmetric way on the path from the parent of leaf $k$ to $w$. The set $V_{jk}$ can be easily found in $O(\log n)$
time by following the two paths from the root to leaf $j$ and leaf
$k$. For each node in $V_{jk}$, we increase its shift value by
$\delta$. This finishes the right-shift process, which can be done in $O(\log n)$ time. Similarly, each left-shift process can also be done in $O(\log n)$ time.

After the algorithm finishes, we can use $T_p$ to
obtain the locations for all sensors in $O(n\log n)$ time.

\subsection{The Left-Shift Tree $T_l$ and the Global Variable $\gamma$}

It remains to compute the size $|S'_l(g_i)|$ and the smallest
displacement of the sensors in $S'_l(g_i)$. Our goal is to compute them in
$O(\log n)$ time. This is one main difficulty in our containing case algorithm.
We propose a {\em left-shift tree} $T_l$ to maintain the displacement
information of the sensors that have positive displacements (i.e.,
their current positions are to the left of their original locations).

The tree $T_l$ is a complete binary tree of $n$ leaves and $O(\log n)$
height. The leaves from left to right correspond to the $n$ sensors. For each leaf $j$,
denote by $\pi_j$ the path in $T_l$ from the root to the leaf.
Each node $v$ of $T_l$ is associated with the following information.

\begin{enumerate}
\item
If $v$ is a leaf, then $v$ is associated with a {\em flag},
and $v.flag$ is set to ``valid'' if
the current displacement of $s_i$ is positive and ``invalid''
otherwise.  Initially all leaves are invalid.
If the flag of leaf $j$ is valid/invalid, we also say the sensor $s_j$ is valid/invalid.
Thus, $S'_l(g_i)$ is the set of valid sensors of $S_l(g_i)$.

\item
As in the position tree $T_p$, regardless of whether $v$ is an internal node or a leaf, $v$ maintains a {\em shift} value $v.shift$.
At any moment during the algorithm, for each leaf $j$, the sum of all
shift values of the nodes in the path $\pi_j$ is exactly the negative value of the current displacement of the sensor $s_j$.

\item

Node $v$ maintains a {\em min} value $v.min$, which is equal to $d$
minus the sum of the shift values of the nodes in the path from $v$ to
the root, where $d$ is
the smallest displacement among all valid leaves in the subtree rooted
at $v$, and further, the index of the corresponding sensor that has
the above smallest displacement $d$ is also maintained in $v$
as $v.index$.

If no leaves in the subtree of $V$ are valid, then $v.min=+\infty$ and $v.index=0$.

\item
Node $v$ maintains a {\em num} value $v.num$, which is the number of valid leaves in the subtree of $v$. Initially $v.num=0$ for all nodes.

\end{enumerate}

The tree $T_l$ can support the following operations in $O(\log n)$ time each.

\begin{description}
\item[set-valid]
Given a sensor $s_j$, the goal of this operation is to set the flag of the $j$-th leaf valid.

To perform this operation, we first find the leaf $j$, denoted by $u$.
We set $u.flag=valid$, $u.min=0$, $u.index=j$. Next, we update the min and index values of the other nodes in the path $\pi_j$ in a bottom-up manner.
Beginning from the parent of $u$, for each
node $v$ in $\pi_j$, we set $v.min = \min\{v_l.min+v_l.shift, v_r.min+v_r.shift\}$ where
$v_l$ and $v_r$ are the left and right child of $v$, respectively, and we set $v.index$ to $v_l.index$ if $v_l$ gives the above minimum value and $v_r.index$ otherwise.

Finally, we update the num values for all nodes in the path $\pi_j$ by increasing $v.num$
by one for each node $v\in \pi_j$.

Hence, the set-valid operation can be done in $O(\log n)$ time.

\item[set-invalid]
Given a sensor $s_j$, the goal of this operation is to set the flag of the $j$-th leaf invalid.

We first find leaf $j$, set it invalid, set its min value to $+\infty$, and set its num value to $0$. Then, we
update the min, index, and num values of the nodes in the path $\pi_j$
similarly as in the above set-valid operation. We omit the details. The set-invalid
operation can be done in $O(\log n)$ time.

\item[left-shift]

Given two indices $j$ and $k$ with $j\leq k$, as well as a distance
$\delta$, the goal of this operation is to move each sensor in $S(j,k)$
leftwards by $\delta$. It is required that $\delta$ is small enough such that any valid (resp., invalid) sensor before the operation is still valid (resp., invalid) after the operation.

The operation can be performed in a similar way as we did on the
position tree $T_p$, with the difference that we also need to update the
shift, min, and index values of some nodes.
Specifically, we first compute the set $V_{jk}$ of $O(\log n)$ nodes,
as defined in the position tree $T_p$, and then for each
node $v$ of $V$, we {\em increase} its shift value by $\delta$.

Next, we update the min and index values. An easy
observation is that only those nodes on the two paths $\pi_j$ and
$\pi_k$ need to have their min and index values updated. Specifically, for
$\pi_j$, we follow it from leaf $j$ in a bottom-up manner, for each
node $v$, we update $v.min$ and $v.index$ in the same way as we did in
the set-valid operations. We do the similar things for the path
$\pi_k$. The time for performing this operation is $O(\log n)$.

\item[right-shift]
Given two indices $j$ and $k$ with $j\leq k$, as well as a distance
$\delta$, the goal of this operation is to move each sensor in $S(j,k)$
rightwards by $\delta$. Similarly, $\delta$ is small enough such that any valid (resp., invalid) sensor before the operation is still valid (resp., invalid) after the operation.

This operation can be performed in a symmetric way as the above
left-shift operation and we omit the details.

\item[find-min]
Given two indices $j$ and $k$ with $j\leq k$,
the goal is to find the smallest displacement and the corresponding
sensor among all valid sensors in $S(j,k)$.

We first find the set $V_{jk}$ of $O(\log n)$ nodes as before.
For each node $v\in V_{jk}$, we compute the smallest
displacement among all valid nodes in its subtree, which is equal to $v.min$ plus
the shift values of the nodes in the path from $v$ to
the root. These smallest displacements for all nodes in $V_{jk}$ can be
computed in $O(\log n)$ time in total by traversing the two paths
$\pi_j$  and $\pi_k$ in the top-down manner. The smallest displacement
among all valid sensors in $S(j,k)$ is the minimum among all above $O(\log
n)$ smallest displacements, and the corresponding
sensor for the smallest displacement can be immediately obtained by using
$v.index$ associated with each node of $V_{jk}$. Thus, each find-min
operation can be done in $O(\log n)$ time.

\item[find-num]
Given two indices $j$ and $k$ with $j\leq k$,
the goal is to find the number of valid sensors in $S(j,k)$.

We first find the set $V_{jk}$ of $O(\log n)$ nodes as before, and then return the
sum of the values $v.num$ for all nodes $v\in V_{jk}$. Hence,
$O(\log n)$ time is sufficient for performing the operation.
\end{description}

In addition, our algorithm maintains a global variable $\gamma$ that is
the rightmost sensor that has ever been moved to the left. We will use
$\gamma$ to determine whether we should do a set-valid operation on a
sensor in the left-shift tree $T_l$ and make sure the total number of set-valid
operations on $T_l$ in the entire algorithm is at most $n$. Initially,
$\gamma=0$. As will be clear later, the variable $\gamma$ will never
decrease during the algorithm.

\subsection{The $O(n\log n)$ Time Algorithm}

Using the three trees $T_o$, $T_p$, $T_l$, and the global variable
$\gamma$, we implement the algorithm \cite{ref:CzyzowiczOn10}
described in Section \ref{sec:pre} in $O(n\log n)$ time, as follows.

The initialization of these trees can be easily done in $O(n\log n)$
time. Suppose the algorithm is about to consider gap $g_i$. We assume the three
trees and $\gamma$ have been correctly maintained. We first use use
the overlap tree $T_o$ to find the two overlaps $o_i^r$ and $o_i^l$ in
$O(\log n)$ time, as discussed earlier. The two numbers
$|S_l(g_i)|$ and $|S_r(g_i)|$, as well as the cost $C(o_i^r)$,
are also determined. Next, we find $|S'_l(g_i)|$ by doing a find-num operation on $T_l$
using the index of the right generator of $o_i^l$ and the index of the
left generator of $g_i$. The cost $C(o_i^r)$ is thus obtained.
Depending on whether $C(o_i^r)< C(o_i^l)$, we have two main cases.

\subsubsection{Case $C(o_i^r)< C(o_i^l)$}
If $C(o_i^r)< C(o_i^l)$,
we do a left-shift process that moves
all sensors in $S_r(g_i)$ leftwards by distance
$\delta=\min\{|g_i|,|o_i^r|\}$. Note that $S_r(g_i)=S(j,k)$ with $j$
being the index of the right generator of $g_i$ and $k$ being the
index of the left generator of $o_i^r$. To implement the above
left-shift process, we first do a left shift on the position tree $T_p$, as described
earlier. Then, we update the left-shift tree $T_l$ and the variable
$\gamma$ in the following way.

Since $o_i^r$ is an overlap and the gaps that have been
covered are all to the left of $o_i^r$, no sensor
to the right of $o_i^r$ has ever been moved. Specifically, sensor $s_t$
has never been moved, for any $t>k$. This implies that
$\gamma\leq k$.

If $\gamma < j$, then for each sensor $s_t$ with $j\leq t\leq k$,
we first do a set-valid operation on $T_l$ on $s_t$ and
then do a left-shift operation on $T_l$ on $s_t$ with distance $\delta$.

If $j\leq \gamma < k$, we have the following lemma.

\begin{lemma}\label{lem:10}
If $j\leq \gamma < k$, then right before the above left-shift process,
all sensors in $S(j,\gamma)$ have positive displacements and thus are valid.
\end{lemma}
\begin{myproof}
We consider the situation right before the above left-shift process.

First of all, we claim that the displacement of $s_\gamma$ must be
positive. Indeed, according to the definition of $\gamma$, if
the displacement of $s_{\gamma}$ is not positive, then there must be
a shift process previously in the algorithm that moved $s_{\gamma}$ rightwards. However,
since the gaps that have been considered by the algorithm are all to the left of
$g_i$ and thus to the left of $s_{\gamma}$, $s_{\gamma}$ never had any
chance to be moved rightwards. The claim thus follows.
Hence, if $j=\gamma$, the lemma is trivially true.

If $j<\gamma$, assume to the contrary that there is a sensor $s_t$ with $j\leq t< \gamma$
whose displacement is not positive. Since the
displacement of $s_{\gamma}$ is positive, the above situation
can only happen if the algorithm covered a gap between $s_t$ and
$s_{\gamma}$, which contradicts with the fact that all gaps that have
been covered by the algorithm are to the left of $g_i$ and thus to the
left of $s_j$. Thus, the lemma follows.
\end{myproof}

If $j\leq \gamma < k$, for each $t$ with $\gamma<t\leq k$, we first do a
set-valid operation on $s_t$ and then do a left-shift operation on $s_t$ with
distance $\delta$ in $T_l$.
Finally, we do a left-shift operation for the sensors in $S(j,\gamma)$
on $T_l$ with distance $\delta$. Based on Lemma \ref{lem:10},
the tree $T_l$ is now correctly updated.

Note that during the above left-shift process,
we did multiple set-valid operations and each of them is followed immediately by a left-shift
operation. An observation is that the total
number of set-valid operations in the entire algorithm is
at most $n$, because the sensors that
are set to valid during this left-shift processes have never been set to valid
before as their indices are larger than
$\gamma$. The number of left-shift operations immediately following these
set-valid operations is thus also at most $n$.

Finally, we update $\gamma$ to $k$.

If $|g_i|<|o_i^r|$, we proceed on the next gap $g_{i+1}$.
Otherwise, $o_i^r$ is eliminated and we delete it from the overlap
tree $T_o$. Since $g_i$ is only partially covered, we proceed on the
remaining $g_i$ with the same approach (in the special case
$|g_i|=|o_i^r|$, we proceed on $g_{i+1}$).

\subsubsection{Case $C(o_i^r)\geq C(o_i^l)$}
If $C(o_i^r)\geq C(o_i^l)$, we perform a right-shift process that
moves all sensors in $S_l(g_i)$ rightwards by
distance $\delta = \min \{|g_i|, |o_i^l|, \alpha\}$, where $\alpha$ is
the smallest displacement of the sensors in $S'_l(g_i)$.
Let $j$ be the index of the right
generator of $o_i^l$ and $k$ be the index of the left generator of
$g_i$. Hence, $S_l(g_i)=S(j,k)$.

To implement the right-shift process, we first do a find-min operation
on $T_l$ with indices $j$ and $k$ to compute $\alpha$. Then,
we update the position tree $T_p$ by doing a right-shift
operation for the sensors in $S(j,k)$ with distance $\delta$. Since no sensor is moved leftwards in the
above process, we do not need to update $\gamma$.

Next, we update the other two trees $T_o$ and $T_l$, depending on
which of the three values $|g_i|$, $|o_i^l|$, and $\alpha$ is the smallest.

If $\delta=\alpha$, we do a right-shift operation with
indices $j$ and $k$ for distance $\delta=\alpha$ on $T_l$. Recall that
the find-min operation can also return the sensor that gives the
sought smallest displacement. Suppose the above find-min operation on $T_l$ returns $s_t$ whose displacement is $\alpha$, with $j\leq t\leq k$.
Since the displacement of $s_t$ now becomes zero,
we do a set-invalid operation on $s_t$ in $T_l$. Note that although it is
possible that $\gamma=t$, we do not need to update $\gamma$.

We should point out a subtle situation where multiple sensors in
$S'_l(g_i)$ had displacements equal to $\alpha$.
For handling this case, we do another
find-min operation on $T_l$ with indices $j$ and $k$. If the smallest
displacement found by the operation is zero, then we do the
set-invalid operation on $T_l$ on the sensor returned by this
find-min operation. We keep doing the find-min operations
until the smallest displacement found above is larger than zero.
Although there may be multiple set-invalid and find-min operations
during the above procedure, the total number of these operations is
$O(n)$ in the entire algorithm. To see this, it is sufficient to
show that the number of set-invalid operations is $O(n)$ because there is exactly one
find-min operation following each set-invalid operation.
After each set-invalid operation, say, on a sensor $s_t$, we claim
that the sensor $s_t$ will never be set to valid again in
the algorithm. Indeed, since the
displacement of $s_t$ was positive, according to the definition of
$\gamma$, we have $t\leq \gamma$. Since each set-valid operation is
only on sensors with indices larger than $\gamma$ and the value
$\gamma$ never decreases, $s_t$ will never be set to valid again in
the algorithm. In fact, $s_t$ will never be moved leftwards
in the algorithm because $s_t$ is to the
left $g_i$ and all gaps that will be covered in the algorithm
are to the right of $g_i$ and thus are to the right of $s_t$.

This finishes the discussion for the case $\alpha=\delta$.
Below we assume $\delta<\alpha$.

We do the right-shift operation with indices $j$ and $k$ for
distance $\delta$ on $T_l$. Since $\delta<\alpha$, no valid sensor
in $S'_l(g_i)$ will become invalid due to the right-shift.
If $\delta=|o_i^l|$, we delete $|o_i^l|$ from $T_o$ since $o_i^l$ is
eliminated. If $\delta=|g_i|$, we proceed on the next gap $g_{i+1}$;
otherwise, we proceed on the remaining $g_i$.

The algorithm finishes after all gaps are covered. The above
discussion also shows that the running time of the algorithm is
bounded by $O(n\log n)$.

\section{The One-Sided Case}
\label{sec:onesided}

In this section, we solve the one-sided case in $O(n\log n)$ time, by
using our algorithm for the containing case in Section
\ref{sec:contain} as an initial step. In the one-sided case,
the sensors whose covering intervals do not intersect $B$ are all
in one side of $B$, and without loss of generality, we assume it is
the right side. Specifically, we assume $0\leq x_1+z$ holds.
We assume at least one sc-interval does not intersect $B$ since
otherwise it would become the containing case.
Note that this implies $\beta < x_n-z$.

We use {\em configuration} to refer to a specification of where each
sensor is located. For example, in the input configuration,
each sensor $s_i$ is located at $x_i$.

A sequence of consecutive sensors $s_i,s_{i+1},\ldots s_j$ are said to be in {\em attached positions} if for each $i\leq k\leq j-1$ the right endpoint of the covering interval $I(s_k)$ of $s_k$ is at the same position as the left endpoint of $I(s_{k+1})$.

\subsection{Observations}

First, we show in the following lemma that a special case where no
sc-interval intersects $B$, i.e., $\beta<x_1-z$, can be easily solved
in $O(n)$ time.

\begin{lemma}\label{lem:20}
If $\beta<x_1-z$, we can find an optimal solution in $O(n)$ time.
\end{lemma}
\begin{myproof}
If $\beta<x_1-z$, then all sensor covering intervals are strictly to the right side of $B$.
According to the order preserving property, in the optimal solution
$I(s_1)$ must have its left endpoint at $0$ (i.e., $s_1$ is at $z$).
Note that we need at least $\lceil\frac{\beta}{2z}\rceil$ sensors to fully
cover $B$. Since all sensors have their covering intervals strictly to the
right side of $B$ and no sensor intersects $B$, in the optimal
solution sensors in $S(1,\lceil\frac{\beta}{2z}\rceil)$ must be in
attached positions. Therefore, the optimal solution has a very special
pattern: $s_1$ is at $z$, sensors in
$S(1,\lceil\frac{\beta}{2z}\rceil)$ are in attached positions, and
other sensors are at their original locations. Hence, we can compute
this optimal solution in $O(n)$ time.
\end{myproof}

In the following, we assume $\beta\geq x_1-z$, i.e.,  $I(s_1)$ intersects $B$.
Let $m$ be the largest index such that $I(s_m)$ intersects $B$.
Note that $m<n$ due to $\beta<x_n-z$.
To simplify the notation, let $S_I=S(1,m)$ and $S_R=S(m+1,n)$.

Our containing case algorithm is not applicable here and one
can easily verify that the cost function we used in the containing
case do not work for the sensors in $S_R$. More specifically,
suppose we want to move a sensor $s_i$ in $S_R$ leftwards to cover a gap;
there will be an ``additive'' cost $x_i-z-\beta$, i.e., $I(s_i)$ has
to move leftwards by that
distance before it touches $B$. Recall that the cost we defined on
overlaps in the containing case is a ``multiplicative'' cost, and the
above additive cost is not consistent with  the
multiplicative cost. To overcome this difficulty, we have to use a
different approach to solve the one-sided case.

Our main idea is to somehow transform the one-sided case to the
containing case so that we can use our containing case algorithm.
Let $D_{opt}$ be any optimal solution for our problem. By slightly
abusing notation, depending on the context, a ``solution'' may either refer to the
configuration of the solution or the
sum of moving distances of all sensors in the solution.
If no sensor of $S_R$ is moved in $D_{opt}$, then
we can compute $D_{opt}$ by running our
containing case algorithm on the sensors in $S_I$.
Otherwise, let $m^*$ be the largest index such that
sensor $s_{m^*}\in S_R$ is moved in $D_{opt}$. If we know
$m^*$, then we can easily compute $D_{opt}$ in $O(n\log n)$ time as
follows. First, we ``manually'' move all sensors in
$S(m+1,m^*)$ leftwards to $\beta+z$ such that the left endpoints of their covering
intervals are at $\beta$. Then, we apply our containing
case algorithm on all sensors in $S_{1m^*}$,
which now all have their
covering intervals intersecting $B$ (which is an instance of the
containing case), and let $D(m^*)$ be the solution obtained above. Based on
the order preserving property, the
following lemma shows that $D(m^*)$  is $D_{opt}$.

\begin{lemma}\label{lem:30}
$D(m^*)$ is $D_{opt}$.
\end{lemma}
\begin{myproof}
Since $s_{m^*}$ is moved in $D_{opt}$, $I(s_{m^*})$ must intersect
$B$ in $D_{opt}$. Based on the order preserving property, for each
$m+1\leq i\leq m^*$, $I(s_i)$ intersects $B$  in $D_{opt}$, which
implies that the location of $s_i$ in $D_{opt}$ must be to the left of
$\beta+z$. On the other hand, since no sensor $s_i$ with $i>m^*$ is
moved, sensors in $S(m^*+1,n)$ are useless for computing $D_{opt}$.
Therefore, $D_{opt}$ is essentially the optimal solution for the
containing case on $S(1,m^*)$ after each sensor in $S(m+1,m^*)$ is
moved leftwards to $\beta+z$, i.e., $D_{opt}=D(m^*)$. The lemma thus follows.
\end{myproof}

By the above discussion, one main task of our algorithm is to determine $m^*$.

For each $j$ with $m<j\leq n$, let $D_s(j)=\sum_{i=m+1}^j (x_i-z-\beta)$,
i.e., the sum of
the moving distances for ``manually'' moving all sensors in $S(m+1,j)$ leftwards
to $\beta+z$, and we use $F_j$ to denote the configuration after
the above manual movement and we let $F_j$ contain only the sensors in $S(1,j)$ (i.e.,
sensors in $S(j+1,n)$ do not exist in $F_j$). Let $D_s(m)=0$ and $F_m$ be the input
configuration but containing only sensors in $S(1,m)$.
For each $m\leq j\leq n$, suppose we apply our
containing case algorithm on $F_j$ and denote
by $D_c(j)$ the solution (in the case where
$\beta>2zj$, we let $D_c(j)=+\infty$), and further, let $D(j)=D_s(j)+D_c(j)$.

The above discussion leads to the following lemma.

\begin{lemma}\label{lem:40}
	$D_{opt}=\min_{m\leq j\leq n}D(j)$ and $m^*=\argmin_{m\leq j\leq n}D(j)$.
\end{lemma}

\subsection{The Algorithm Description and Correctness}

Lemma \ref{lem:40} leads to a straightforward $O(n^2\log n)$ time
algorithm for the one-sided case, by computing
$D(j)$ in $O(n\log n)$ time for
each $j$ with $m\leq j\leq n$, as suggested above.
In the sequel, by exploring the solution structures, we present an
$O(n\log n)$ time solution. The algorithm itself is simple, but it is
not trivial to discover the observations behind the scene.


Our algorithm will compute $D(j)$ for all $j=m,m+1,\ldots,n$.
Recall that $D(j)=D_s(j)+D_c(j)$. Since it
is easy to compute all $D_s(j)$'s in $O(n)$ time, we focus on computing
$D_c(j)$'s. The main idea is the following. Suppose we already have the solution
$D_c(j-1)$, which can be considered as being obtained by our
containing case algorithm. To compute $D_c(j)$, since we have an additional
overlap defined by $s_j$ at $\beta+z$, i.e., the sc-interval $I(s_j)$, we modify
$D_c(j-1)$ by ``reversing'' some shift processes that have been
performed in the containing algorithm when
computing $D_c(j-1)$, i.e., using $I(s_j)$ to cover some gaps that
were covered by other overlaps in $D_c(j-1)$. The details are given
below.

We first compute $D_c(m)$ on the configuration $F_m$. If $2zm<\beta$, then  $D_c(j)=+\infty$ for each $m\leq j< \lceil\frac{\beta}{2z}\rceil$; in this case, we can start from computing $D_c(\lceil\frac{\beta}{2z}\rceil)$ and use the similar idea as the following algorithm. To make it more general, we
assume $m\geq \lceil\frac{\beta}{2z}\rceil$, and thus $D_c(m)<+\infty$.

Consider our containing case algorithm
for computing $D_c(m)$.
Recall that our containing case algorithm consists of shift processes
and each shift process covers a gap using an overlap.  Let $p_1,p_2,\ldots,p_q$ be
the shift processes performed in the algorithm in the {\em inverse} order of time
(e.g., $p_1$ is the last process), where $q$ is the total number of processes in the algorithm.
For each $1\leq i\leq q$, let $g_i$ be the gap covered in the process $p_i$
by using/eliminating an overlap, denoted by $o_i$. Note that each gap/overlap above may
not be an original gap/overlap in the input configuration but only a subset of
an original gap/overlap. It holds that $|o_i|=|g_i|$ for each $1\leq
i\leq q$. We call $G=\{g_1,g_2,\ldots,g_q\}$ the {\em gap list} of $D_c(m)$.
For each $i$, we use $C(o_i)$ to denote the cost
of $o_i$ when the algorithm uses $o_i$ to cover $g_i$ in the process $p_i$. Note that the
above {\em  process information} can be explicitly stored during our containing case
algorithm without affecting the overall running time asymptotically.
We will use these information later.
Note that according to our algorithm the
gaps in $G$ are sorted from right to left.

Next, we compute $D_c(m+1)$, by modifying the configuration $D_c(m)$.
Comparing with $F_m$, the configuration $F_{m+1}$ has an additional
overlap defined by $s_{m+1}$ at $\beta+z$, and we use $o(s_{m+1})$ to denote it.
We have the following lemma.

\begin{lemma} \label{lem:50}
$D_c(m+1)=D_c(m)$ holds if one of the following happens: (1) the coordinate of
the right endpoint of $I(s_m)$ is strictly larger than $\beta$; (2)
$o_1$ is to the right of $g_1$; (3) $o_1$ is to the left of $g_1$ and
the cost $C(o_1)$ is not greater than the number of sensors between $g_1$ and $s_{m+1}$.
\end{lemma}
\begin{myproof}
We prove Case (3) first.

Suppose that we run our containing case algorithm on
both $F_m$  and $F_{m+1}$ simultaneously. We use $A_m$ to denote the algorithm on $F_m$ and use $A_{m+1}$ to denote the algorithm on $F_{m+1}$.
Below we will show that every
shift process of $A_m$ and $A_{m+1}$ is exactly the same, which proves $D_c(m+1)=D_c(m)$.

Consider any shift process $p_j$.
We assume the processes before $p_j$ on both algorithms are the same,
which holds for $j=q$. In $A_m$, the process covers gap $g_j$
by using overlap $o_j$.

If $o_j$ is to the right of $g_j$, then since $o(s_{m+1})$ is the
rightmost overlap in $F_{m+1}$, algorithm $A_{m+1}$ also uses
$o_j$ to cover $g_j$, which is the same as $A_m$.

If $o_j$ is to the left of $g_j$, then depending on whether
$o(s_{m+1})$ is the only overlap to the right of $g_j$, there are two
cases.
\begin{enumerate}
\item
If $o(s_{m+1})$ is not the only overlap to the right $g_j$, then let
$o$ be the closest overlap to $g_j$ among the overlaps to the right of $g_j$. According
to our containing algorithm, the current process of the algorithm only
depends on the costs of the two overlaps $o_j$ and $o$. Hence, algorithm $A_{m+1}$ uses the same shift process to cover $g_j$ as
that in $A_m$, i.e, use $o_j$ to cover $g_j$.

\item
If $o(s_{m+1})$ is the only overlap to the right $g_j$, then the current process of algorithm $A_{m+1}$ depends on the costs of the
two overlaps $o_j$ and $o(s_{m+1})$. In the following, we show that
$C(o_j)\leq C(o(s_{m+1}))$, and thus algorithm $A_{m+1}$ also uses $o_j$ to cover $g_j$, as in $A_m$.

Recall that the list of gaps $g_1,g_2,\ldots,g_q$ are sorted from
right to left by their generators. Thus, the gaps
$g_j,g_{j-1},\ldots,g_1$ are sorted from left to right.
Since $o(s_{m+1})$ is the only overlap to the right $g_j$ in $A_{m+1}$,
there is no overlap in $A_m$
to the right of $g_t$ for any $t$ with $j\geq t\geq 1$.
Hence, algorithm $A_m$ will have to uses the overlaps to the left of $g_t$ to cover
$g_t$ for each $t$ with $j\geq t\geq 1$. In other words, all overlaps
$o_j,o_{j-1},\ldots,o_1$ are to the left of all gaps
$g_j,g_{j-1},\ldots,g_1$, which implies that the above list of
overlaps are sorted from right to left and $C(o_j)\leq C(o_{j-1})\leq
\cdots\leq C(o_1)$.

Since $g_1$ is to the right of $g_j$, the cost $C(o(s_{m+1}))$, which
is the number of sensors between $g_j$ and $s_{m+1}$, is no less than
the number of sensors between $g_1$ and $s_{m+1}$. Since in Case (3)
the number of sensors between $g_1$ and $s_{m+1}$ are at least $C(o_1)$, we obtain that $C(o(s_{m+1}))\geq C(o_1)\geq
C(o_j)$.
\end{enumerate}

The above shows that every shift process of $A_m$ and $A_{m+1}$ is the same, which proves that $D_c(m+1)=D_c(m)$ holds for Case (3).

The proofs of the first two cases are similar to the above, and we only sketch them below.

Case (1) means that sensor $s_m$ still defines an overlap, say $o$, to the right
of $B$. If we run our containing case algorithm on $F_{m+1}$ to
compute $D_c(m+1)$, sensor $s_{m+1}$ will not be moved since the
overlap $o(s_{m+1})$ is to the right of $o$. Hence, $D_c(m+1)=D_c(m)$
holds.

Case (2) means the last shift process covers $g_1$ using $o_1$ that is
to the right of $g_1$. If we
run our containing case algorithm on $F_{m+1}$, overlap $o(s_{m+1})$
will never have any chance to be used to cover any gap, because $o(s_{m+1})$ is the
rightmost overlap of $F_{m+1}$. Hence, $D_c(m+1)=D_c(m)$
holds.
\end{myproof}

To compute $D_c(m+1)$, we first check whether one of the three cases in Lemma \ref{lem:50}
happens, which can be
done in constant time by the above process information stored when computing $D_c(m)$.
If any of the three cases happens, we are done for computing
$D_c(m+1)$. Below, we assume none of the cases happens.

Let $C(s_{m+1},g_1)$ be the number of sensors between $g_1$ and
$s_{m+1}$, which would be the cost of the overlap $o(s_{m+1})$ if it
were there right before we cover $g_1$. Note that since we know the
generators of $g_1$, $C(s_{m+1},g_1)$ can be computed in constant time
(e.g., if $g_1$ has two generators, $C(s_{m+1},g_1)=m+1-a+1$, where
$a$ is the index of the right generator of $g_1$).

Define $R(g_1)$ to be $C(s_{m+1},g_1)-C(o_1)$.
We can consider $R(g_1)$ as the
``unit revenue'' (or savings) if we use $o(s_{m+1})$ to cover $g_1$ instead of
using $o_1$. Note that $R(g_1)>0$ otherwise the third case of Lemma
\ref{lem:50} would happen. Hence, it is possible to obtain a
better solution than $D_c(m)$ by using $o(s_{m+1})$ to cover $g_1$
instead of $o_1$. Note that $|g_1|\leq 2z$ and $|o(s_{m+1})|=2z$.

If $|o(s_{m+1})|=|g_1|$, then we use $o(s_{m+1})$ to cover $g_1$.
Specifically, we move all sensors in $S(a,m+1)$ leftwards
by distance $|g_1|$, where $a$ is the index of the right generator of
the overlap $o_1$.
The above essentially ``restores'' the overlap $o_1$ and covers $g_1$
by eliminating $o(s_{m+1})$. We refer to it as a {\em reverse operation}
(i.e., it reverses the shift process that covers $g_1$ by using $o_1$ in the algorithm for computing $D_c(m)$). Due to $|o(s_{m+1})|=|g_1|$, after the reverse operation, $g_1$ is fully covered by $o(s_{m+1})$ and $o(s_{m+1})$ is eliminated.
We will show later in Lemma \ref{lem:60} that the current configuration is $D_c(m+1)$.
Note that $D_c(m+1)=D_c(m)-R(g_1)\cdot |g_1|$.
Again, $o_1$ is restored in $D_c(m+1)$. Finally, we  remove $g_1$ from the list $G$.

If $|g_1|<|o(s_{m+1})|$, then we do a revere operation by using $o(s_{m+1})$ to cover $g_1$ and restore $o_1$,
after which $o(s_{m+1})$ is not eliminated but becomes shorter. We remove $g_1$ from $G$ and proceed on the next gap
$g_2$.

In general, suppose we have covered gaps $g_1,g_2,\ldots,g_k$
by using $o(s_{m+1})$ and the overlap $o(s_{m+1})$ still partially
remains (i.e., $\sum_{t=1}^k|g_i|<2z$). The above gaps have all been removed from $G$.
Let $F'$ denote the current configuration. If $G$ is now empty, then
we are done with computing $D_c(m+1)$, which is equal to
$D_c(m)-\sum_{t=1}^kR(g_t)\cdot |g_t|$; otherwise,
we consider gap $g_{k+1}$, as follows.

Similar to Lemma \ref{lem:50}, we will show later in Lemma
\ref{lem:60} that $F'$ is $D_c(m+1)$ if
one of the following two cases happens: (1) $o_{k+1}$ is to the right
of $g_{k+1}$; (2) $o_{k+1}$ is to the left of $g_{k+1}$ but
$C(o_{k+1})$ is not greater than the number of sensors between $g_{k+1}$
and $s_{m+1}$. If one of the above two cases happens, then we are done
with computing $D_c(m+1)$, which is equal to
$D_c(m)-\sum_{t=1}^kR(g_t)\cdot |g_t|$. Otherwise, we do the
following. Note that the length of $o(s_{m+1})$ in $F'$ is
$2z-\sum_{t=1}^k|g_t|$. Depending on whether $|o(s_{m+1})|\geq
|g_{k+1}|$, there are two cases. As for $g_1$, we define $C(s_{m+1},
g_{k+1})$ as the number of sensors between $g_{k+1}$ and $s_{m+1}$, and
define $R(g_{k+1})=C(o_{k+1})-C(s_{m+1}, g_{k+1})$.

\begin{enumerate}
\item
If $|o(s_{m+1})|\geq |g_{k+1}|$, then we do a reverse operation to
cover $g_{k+1}$ by using $o(s_{m+1})$.
If $|o(s_{m+1})|= |g_{k+1}|$, we are done with computing $D_c(m+1)$,
which is equal to $D_c(m)-\sum_{t=1}^{k+1}R(g_t)\cdot |g_t|$;
otherwise, we proceed on the next gap $g_{k+2}$. In either case, we
remove $g_{k+1}$ from $G$, and the reverse operation restores the
overlap $o_{k+1}$.

\item
If $|o(s_{m+1})|< |g_{k+1}|$, then $o(s_{m+1})$ is not long enough to
cover $g_{k+1}$. We do a reverse operation to use $o(s_{m+1})$ to
partially cover $g_{k+1}$ of length $|o(s_{m+1})|$, and the remaining
part of $g_{k+1}$ is still covered by $o_{k+1}$. We are done with
computing $D_c(m+1)$, which is equal to
$D_c(m)-\sum_{t=1}^{k}R(g_t)\cdot |g_t| - R(g_{k+1})\cdot
|o(s_{m+1})|$. Since $g_{k+1}$ still partially remains in $D_c(m+1)$,
we do not remove $g_{k+1}$ from $G$ but change its size accordingly.
In addition, overlap $o_{k+1}$ is partially restored in $D_c(m+1)$ because
its size is $|o(s_{m+1})|$, which is smaller than its original size.
\end{enumerate}

The algorithm stops after $D_c(m+1)$ is obtained.

\begin{lemma}\label{lem:60}
The solution obtained in the above algorithm is $D_c(m+1)$.
\end{lemma}
\begin{myproof}
Let $F$ be the configuration obtained by our algorithm.
Below we show that $F$ is $D_c(m+1)$.
If one of the three cases in Lemma \ref{lem:50} happens, then by Lemma \ref{lem:50}, $F$ is $D_c(m+1)$. Below we
 assume none of the three cases in Lemma \ref{lem:50} happens.

Suppose that we run our containing case algorithm on
both $F_m$ and $F_{m+1}$ simultaneously. Let $A_m$ be the algorithm on $F_m$ and let $A_{m+1}$ be the algorithm on $F_{m+1}$.

Consider any shift process $p_i$ of $A_m$. We assume the processes before $p_i$ on both algorithms are the same, which holds for $i=q$.
By the proof of Lemma \ref{lem:50},
$p_i$ may not be the same in $A_m$ and $A_{m+1}$ only if for each
process $p_j$ after $p_i$ (i.e., $j\leq i$ since the order of
processes follows the reverse order the time),
$o_j$ is to the left of $g_j$, i.e., $p_j$ is a right-shift process.
Therefore, we only need to consider the right-shift
processes after the last left-shift process in $A_m$.

Let $k$ be the smallest index with $0\leq k\leq q-1$ such that $o_{k+1}$
is to the right of $g_{k+1}$, i.e., $p_1,p_2,\ldots,p_k$ are
the right-shift processes after the last left-shift process in $A_m$.
Note that $k\neq 0$ since otherwise Case (2) of Lemma \ref{lem:50}
would happen.
Hence, the process $p_i$ with $k+1\leq i\leq q$ is the same in both $A_m$ and $A_{m+1}$.

Since none of the cases in Lemma \ref{lem:50} happens,
$C(o_1)> C(s_{m+1},g_1)$. Let $t$ be the largest index such that
$C(o_t)>C(s_{m+1}, g_t)$. For each process $p_i$ with $k\leq i\leq
t+1$, since $C(o_i)\leq C(s_{m+1}, g_i)$, the process is the same in
both $A_m$ and $A_{m+1}$. In summary, the above shows that if $q\geq i\geq t+1$,
the process $p_i$ is the same in both $A_m$ and $A_{m+1}$, i.e., the
first $(q-t+1)$ processes in both $A_m$ and $A_{m+1}$ are the same.

Consider the next process $p_t$, which covers gap $g_t$ by using $o_t$ in $A_m$. In $A_{m+1}$, however, using $o(s_{m+1})$ to cover it can
give a better solution. Depending on whether $\sum_{i=1}^t|g_i|\leq 2z$, there are two cases.

\begin{enumerate}
\item
If $\sum_{i=1}^t|g_i|\leq 2z = |o(s_{m+1})|$, then since $o(s_{m+1})$ is long enough,
$A_{m+1}$ will use $o(s_{m+1})$ to cover all gaps from $g_t$ to $g_1$ and thus obtain $D_c(m+1)$.
Let $s_h$ be the left generator of $g_t$ and let $x(s_h)$ be the location of $s_h$ in the configuration right after the process $p_{t+1}$. Since the algorithm $A_{m+1}$ uses $o(s_{m+1})$ to cover all gaps
from $g_t$ to $g_1$, the location of $s_h$ does not change, which implies that the locations of $s_h$ in both $D_c(m+1)$ and $D_c(m)$ are the same, i.e., $x(s_h)$. Similarly, each sensor in $S(1,h)$ has the same location in both $D_c(m+1)$ and $D_c(m)$.
Further, since $o(s_{m+1})$ is the only overlap to the right of $g_t$, all sensors in $S(h,m+1)$ are in attached positions in $D_c(m+1)$.

Now consider the configuration $F$ obtained by our algorithm using the reverse operations. According to our algorithm, only gaps from $g_1$ to $g_t$ will be
covered by $o(s_{m+1})$. Hence, the left generator $s_h$ of $g_t$ does not change
its location. In other words, the position of $s_h$ is the same as that in $D_c(m)$, which is $x(s_h)$. Also, each sensor
in $S(1,h)$ has the same location in both $D_c(m)$ and $F$.
On the other hand, since gaps from $g_1$ to $g_t$ are covered
by $o(s_{m+1})$, the sensors in $S(h,m+1)$ are in attached
positions in $F$.

The above discussion shows that each sensor of $S(1,m+1)$ has the same location
in both $F$ and $D_c(m+1)$, which implies that
$F$ is $D_c(m+1)$.

\item
If $\sum_{i=1}^t|g_i|> 2z$, then $o(s_{m+1})$ is not long enough to cover all
gaps from $g_1$ to $g_t$. Algorithm $A_{m+1}$ will use $o(s_{m+1})$ to cover these gaps in the order
$g_t,g_{t-1},\ldots$ until $o(s_{m+1})$ is eliminated (i.e., $s_{m+1}$ is at
$\beta-z$). Consider the configuration right before $A_{m+1}$
covers $g_t$.  Recall that the algorithm $A_m$ uses
the gaps $o_1,o_2,\ldots,o_t$ to cover all these gaps. Let
$d=\sum_{i=1}^t|g_i|$, which is $\sum_{i=1}^t|o_i|$.

In $A_{m+1}$, according to our discussion above, $o(s_{m+1})$ will
be used first to cover these overlaps for a total length of $2z$, and
then the above overlaps (in the order from right to left)
will be used to cover the remaining
gaps, whose total length is $d- 2z$. Let $h$ be the smallest index such that
$\sum_{i=h}^t|o_i|\geq d-2z$. Then, $A_{m+1}$ will use the overlaps
from $o_t$ to $o_h$ to cover the remaining of the above gaps. Hence,
for each gap $o_i$,
if $h+1\leq i\leq t$, then $o_i$ does not exist in $D_c(m+1)$; if $i\leq h-1$, then
$o_i$ exists there; if $i=h$, then if $\sum_{i=h}^t|o_i|=d-2z$, $o_h$
does not exist, otherwise $o_h$ still exits but become shorter. Let $o_h^1$ be the
subset of $o_h$ that is eliminated and let $o_h^2$ be the rest of
$o_h$ that still exists in $D_c(m+1)$.
Thus, $|o^1_h|+\sum_{i=h+1}^t|o_i|=d-2z$.
Due to $d=\sum_{i=1}^t|o_i|$, it holds  that
$|o_h^2|+\sum_{i=1}^{h-1}|o_i|=2z$.

Now consider the configuration $F$ obtained by our algorithm using reverse operations.
Since $C(o_i)>C(s_{m+1},g_i)$ for each $1\leq i\leq t$, the gaps
$g_1,g_2,\ldots$ will be covered in this order until
$o(s_{m+1})$ is eliminated, implying that
overlaps in $o_1,o_2,\ldots$ will be restored in this order until
$o(s_{m+1})$ is eliminated. Due to $|o_h^2|+\sum_{i=1}^{h-1}|o_i|=2z$, $o_i$
exists in $F$ for each $1\leq i\leq h-1$ and $o_h$ is partially
restored to $o^2_h$ in $F$.

Therefore, in both configurations $F$  and $D_c(m+1)$, overlaps of
$o_1,o_2,\ldots,o_{h-1}$ exist and $o_h$ partially exits as $o_h^2$.
Hence, the two configurations are exactly the same.
\end{enumerate}

The lemma thus follows.
\end{myproof}

Lemma \ref{lem:60} shows that $D_c(m+1)$ is computed correctly. Next,
we use the same approach to compute $D_c(m+2)$ by using the
remaining gaps in $G$. Let $G_m$ denote the remaining $G$.
In order to correctly compute $D_c(m+2)$, one may wonder that
we should use the corresponding gap list of
$D_c(m+1)$ (i.e., the gap list of the containing case algorithm if we apply it on $F_{m+1}$ to compute $D_c(m+1)$), which may not be the same as $G_m$. However, we prove in
Lemma \ref{lem:70} that the result obtained using $G_m$ is $D_c(m+2)$, and further,
this can be generalized to the next solution until $D_c(n)$, i.e.,
we can use the same approach to compute
$D_c(m+3),D_c(m+4),\ldots,D_c(n)$ by using the remaining gaps.

\begin{lemma}\label{lem:70}
If we do the reverse operations on $D_c(m+1)$ and sensor $s_{m+2}$
by using the gaps in $G_m$, then the solution obtained is
$D_c(m+2)$. Similarly, this can be generalized to the next solution
$D_c(m+3)$ and so on until $D_c(n)$.
\end{lemma}
\begin{myproof}
Suppose we apply our containing case
algorithm on the configuration $F_{m+1}$ to compute $D_c(m+1)$ and let
$G'$ be the list of gaps covered in the right-shift processes after
the last left-shift process of the algorithm. Then, using $G'$, we
can compute $D_c(m+2)$ by doing reverse operations on $D_c(m+1)$ and
sensor $s_{m+2}$, and the correctness can be proved similarly to Lemma
\ref{lem:60}. Hence, if $G_m$ is exactly the same as $G'$, then the
lemma trivially follows. However, $G'$ may be the same as $G_m$, as
shown below.

We follow the notations defined in the proof of Lemma \ref{lem:60}. Let
$A_{m+1}$ be our containing case algorithm on $F_{m+1}$ above.
Consider the gap list $G=\{g_1,g_2,\ldots,g_k\}$ for our containing case algorithm on $F_m$.
Again, $G$ only contains the gaps in the right-shift processes after
the last left-shift process. Let $t$ be the largest index such that
$C(o_t)>C(s_{m+1},g_t)$. As in the proof of Lemma \ref{lem:60},
depending on whether $\sum_{i=1}^t|g_i|\leq 2z$, there are two cases.

\begin{enumerate}
\item
If $\sum_{i=1}^t|g_i|\leq 2z$, then as analyzed in Lemma
\ref{lem:60}, algorithm $A_{m+1}$ will use the
overlap $o(s_{k+1})$ to cover all gaps $g_1,g_2,\ldots,g_t$, after
which  the solution $D_c(m+1)$ is obtained. Therefore, the last shift
process of $A_{m+1}$ is a right-shift process, implying that
$G'=\emptyset$. Thus, if we do the reverse operations on
$D_c(m+1)$ and $s_{m+2}$, according to our algorithm,
it holds that $D_c(m+2)=D_c(m+1)$ (similar to Case (2) of Lemma \ref{lem:50}).

On the other hand, the gap list of $G_m$ is $\{g_{t+1},g_{t+2},\ldots,
g_k\}$.

If $\sum_{i=1}^t|g_i|< 2z$, then
the coordinate of the right endpoint of $I(s_{m+1})$ is
strictly larger than $\beta$ in $D_c(m+1)$. According to our reverse operation
algorithm (Case (1) of Lemma \ref{lem:50}), we obtain
$D_c(m+2)=D_c(m+1)$.

Otherwise, the right endpoint of  of $I(s_{m+1})$ is exactly at
$\beta$ in $D_c(m+1)$. We claim that $C(o_{t+1})<C(s_{m+2},g_{t+1})$. To see this,
by the definition of $t$, it holds that $C(o_{t+1})\leq C(s_{m+1},g_{t+1})$.
Note that $C(s_{m+2},g_{t+1})=C(s_{m+1},g_{t+1})+1$, because the right
endpoint of $I(s_{m+1})$ is exactly at $\beta$. Therefore,
$C(o_{t+1})<C(s_{m+2},g_{t+1})$. According to our reverse operation
algorithm (Case (3) of Lemma \ref{lem:50}), we obtain
$D_c(m+2)=D_c(m+1)$.

Therefore, in this case, the solution obtained
by our algorithm is $D_c(m+2)$.

\item
If $\sum_{i=1}^t|g_i|> 2z$, then as analyzed in Lemma
\ref{lem:60}, algorithm $A_{m+1}$ will use $o(s_{k+1})$ to cover gaps
$g_1,g_2,\ldots,g_{h-1}$ and $g_h^1$. Therefore, the list $G'$ is
$\{g_h^2,g_{h+1},\ldots,g_t\}$.

On the other hand, the gap list of $G_m$ is
$\{g_h^2,g_{h+1},\ldots,g_t,g_{t+1},\ldots, g_k\}$,
which is $G'\cup \{g_{t+1},\ldots,g_k\}$.

Note that in this case, the right endpoint of  of $I(s_{m+1})$ is exactly at
$\beta$ in $D_c(m+1)$.

We claim that if we do reverse operations on $D_c(m+1)$ and sensor
$s_{m+2}$, we can obtain the same result using either $G_m$ or $G'$.
Intuitively, due to $C(o_{t+1})<C(s_{m+2},g_{t+1})$, which has been
proved in the above first case, the gaps in
$\{g_{t+1},g_{t+2},\ldots,g_k\}$ are useless for computing $D_c(m+2)$.
The detailed proof for the claim is given below.

Indeed, the reverse operations consider the gaps one by one from the
first gap $g_h^2$. The result can be different only if all gaps of
$\{g_h^2,g_{h+1},\ldots,g_t\}$ are covered by $o(s_{m+2})$, i.e., the
overlap defined by $s_{m+2}$, and $o(s_{m+2})$ has not been fully
eliminated yet.  If this happens, for $G'$, it now becomes empty and thus
according to our reverse operation algorithm the current configuration is $D_c(m+2)$.
For $G_m$, the next gap $g_{t+1}$ is considered.
Due to $C(o_{t+1})<C(s_{m+2},g_{t+1})$, according to our reverse operation
algorithm,  the current solution is $D_c(m+2)$.

Therefore, in this case, the solution obtained
by our algorithm is $D_c(m+2)$.
\end{enumerate}

The above proves that we can compute $D_c(m+2)$ by applying the reverse
operations on $D_c(m+1)$ and $s_{m+2}$ with $G_m$.
Using similar arguments, we can keep computing $D_c(m+3)$ and so on
until $D_c(n)$, by using the remaining gaps in $G$. The lemma thus follows.
\end{myproof}

\subsection{The Algorithm Implementation}

Our algorithm can be easily implemented in $O(n\log n)$ time to
compute the solutions $D_c(i)$ for all $i=m,m+1,\ldots, n$. First, we
can compute $D_c(m)$ in $O(n\log n)$ time by using our containing case
algorithm. During the algorithm, we explicitly record the information
of each shift process $p_i$, as discussed earlier.
In fact, as shown in the
proofs of Lemmas \ref{lem:50} and \ref{lem:60}, we only need to record
all right-shift processes after the {\em last} left-shift process of the
algorithm, and let $G$ be the gap list for the above
right-shift processes (i.e., for each gap $g_i$ in $G$, $o_i$ is to the left of $g_i$).

Next, we apply the reverse operations on $G$ to compute
solutions $D_c(j)$ for $m+1\leq j\leq n$ one by one. To this end, we only need to use the position
tree $T_p$ (the other two trees are not necessary). Each reverse
operation can be done in $O(\log n)$ time using $T_p$ because the
operation essentially moves a sequence of consecutive sensors leftwards
by the same distance. If $G$ becomes $\emptyset$ at any moment during the
algorithm, then the current configuration is the solution we seek. The overall time for computing all solutions $D_c(j)$ for $m+1\leq j\leq n$ is $O(K\cdot \log n)$, where $K$ is the total number of
reverse operations in the entire algorithm. Note that
each reverse operation either covers completely a gap of $G$
or eliminates an overlap $o(s_j)$ for $m+1\leq j\leq n$.
Therefore, $K\leq |G| + n -m=O(n)$.

In summary, we can compute the solutions $D_c(j)$ for all $m\leq j\leq n$
in $O(n\log n)$ time, after which the value $D(j)$ for
all $j=m,m+1,\ldots,n$ as well as the index $m^*$
can be obtained in additional linear time.
Thus, the one-sided case is solved in $O(n\log n)$ time.

\subsection{A Unimodal Property of the Solutions $D(j)$'s}

If there is more than one index $j\in [m,n]$ such that $D(j)=D_{opt}$, then we let
$m^*$ refer to the smallest such index.
The following lemma, which will be useful in Section
\ref{sec:general} for solving the general case, shows a unimodal property of the values $D(j)$ for $j=m,m+1,\ldots,n$.


\begin{lemma}\label{lem:80}
As $j$ increases from $m$ to $n$, the value $D(j)$ first strictly decreases until
$D(m^*)$ and then strictly increases except that
$D(m^*)=D(m^*+1)$ may be possible. Formally,
$D(j-1)>D(j)$ for any $m< j \leq m^*$; $D(m^*)\leq D(m^*+1)$;
$D(j-1)<D(j)$ for any $m^*+2< j\leq n$.
\end{lemma}
\begin{myproof}
To avoid tedious discussion, we make a general position assumption
that no two sensors are at the same position in the input configuration.

Consider any index $j$ with $m<j\leq n$. We have the following.
\begin{equation*}
\begin{split}
	D(j)-D(j-1)
	& = [D_s(j)+D_c(j)]-[D_s(j-1)+D_c(j-1)]\\
	& = [D_s(j)-D_s(j-1)]+[D_c(j)-D_c(j-1)]\\
	& = (x_j-z-\beta)+[D_c(j)-D_c(j-1)].\\
\end{split}
\end{equation*}

Define $f(j)=D_c(j)-D_c(j-1)$. We have the following claim.

{\bf Claim:} {\em $f(j)\leq 0$ and $f(j)$ is nondecreasing as $j$
increases}.

In the sequel, we first prove the lemma by using the above claim and
then prove the claim.

As $j$ increases, since $x_j-z-\beta$ is strictly increasing and
$f(j)$ is nondecreasing, $D(j)-D(j-1)$ is strictly increasing. If
$j=m^*$, then according to the definition of $m^*$,
we have $D(m^*)-D(m^*-1)< 0$. Hence, when $j\leq m^*$,
$D(j)-D(j-1)<0$. On the other hand,  we have
$D(m^*+1)-D(m^*)\geq 0$, and thus, when $j>m^*+1$, $D(j)-D(j-1)>0$.
The lemma thus follows.

In the sequel, we prove the above claim.

We first prove $f(j)\leq 0$. Recall that $D_c(j)$ is the solution
obtained on configuration $F_{j}$ and $D_c(j-1)$ is the solution
obtained on configuration $F_{j-1}$. Comparing with $F_{j-1}$,
$F_j$ has an additional overlap defined by $s_j$ at $\beta+z$, and thus, it holds that
$D_c(j)\leq D_c(j-1)$.  Hence, $f(j)\leq 0$.

Next, we show $f(j)\leq f(j+1)$. Let $|f(j)|$ and $|f(j+1)|$  be the
absolute values of $f(j)$ and $f(j+1)$, respectively. Below we prove $|f(j)|\geq |f(j+1)|$.

Comparing $D_c(j-1)$ with $D_c(j)$, we may consider
the value $|f(j)|$ as the ``marginal revenue'' after having one more
overlap defined by sensor $s_j$ at $\beta+z$.
Intuitively, if we have more sensors, the marginal
revenue will become less and less, i.e., as $j$ increases, $|f(j)|$ is
monotonically decreasing, and thus $|f(j)|\geq |f(j+1)|$.
A detailed proof is given below, which may be skipped if the reader
is confident in the above intuition.

Let $G=\{g_1,g_2,\ldots,g_k\}$ be the gap list of the right-shift
processes after the last left-shift process of our containing
algorithm on $D_c(m)$. We assume the list in $G$ are sorted by the
inverse time order, e.g., $g_1$ is the gap covered by the last
process. Let $O=\{o_1,o_2,\ldots,o_k\}$ be the corresponding overlap
list, and for each $1\leq i\leq h$, let $C(o_i)$ be the cost of $o_i$
during the containing case algorithm.
As analyzed in the proofs of Lemmas \ref{lem:50} and \ref{lem:60}, it holds that
$C(o_1)\geq C(o_2)\geq\cdots\geq C(o_k)$.

Recall that we obtain all solutions $D_c(i)$ for $m+1\leq i\leq
n$ by doing the reverse operations with $G$. More specifically, let
$G_{j-1}$ be the list of remaining gaps of $G$ right after $D_c(j-1)$ is
obtained. To compute $D_c(j)$, we do reverse operations on $D_c(j-1)$
with $s_j$ and $G_{j-1}$.
Let $G_{j-1}=\{g_h,g_{h+1},\ldots,g_k\}$.
Let $G_j$ be the gap list right
after we obtain $D_c(j)$. According
to our algorithm, $G_j$ is obtained from $G_{j-1}$ by removing the
first several gaps that are covered by
$o(s_{j})$, i.e., the overlap defined by $s_j$ at $\beta+z$.
We assume $G_j=\{g_{t},g_{t+1},\ldots,g_k\}$ and thus, $o(s_j)$
has covered the gaps ${g_h,g_{h+1},\ldots,g_{t-1}}$ completely.
Note that depending on
whether $o(s_j)$ is used to cover $g_t$ partially in $D_c(j-1)$, the
$g_t$ in $G_j$ may only be a subset of the $g_t$ in $G_{j-1}$ (i.e.,
they have the same generators but their lengths are different). For
simplicity of discussion, we assume $o(s_j)$ does not partially cover
$g_t$.

Our algorithm computes $D_c(j+1)$ by doing
reverse operations on $D_c(j)$ with sensor $s_{j+1}$ and $G_{j}$.

If the coordinate of the right endpoint of $I(s_j)$ is strictly larger
than $\beta$
in the configuration $D_c(j)$, then according to our algorithm (Lemma
\ref{lem:50}), we have $D_c(j+1)=D_c(j)$.
Hence, $f(i+1)=0$, implying that
$|f(i)|\geq |f(i+1)|$.

In the following, we assume the coordinate of the right endpoint of $I(s_j)$ is
no greater than $\beta$, and thus it is exactly $\beta$ since $s_j$
is the rightmost sensor in the configuration $F_j$. In this case, the
total length of the gaps of $G_{j-1}$ covered by the overlap
$o(s_{j})$ is $2z$. Recall that the gaps of $G_{j-1}$ covered by
$o(s_{j})$ in $D_c(j)$ are $g_h,g_{h+1},\ldots,g_{t-1}$. Thus, we have
$\sum_{i=h}^{t-1}|g_i|=2z$. According to our algorithm, it holds that
$D_c(j)=D_c(j-1)-\sum_{i=h}^{t-1}R(g_i)\cdot|g_i|$, where
$R(g_i)=C(o_i)-C(s_j,g_i)$. Therefore,
$|f(i)|=\sum_{i=h}^{t-1}R(g_i)\cdot|g_i|$. Since gaps in $G_{j-1}$ are
sorted from right to left,
an easy observation is that $C(s_j,g_i)\geq C(s_j,g_{i-1})$ for any
$h+1\leq i\leq t-1$, i.e., as $i$ increases, $C(s_j,g_i)$ increases.
Recall that as $i$ increases, $C(o_i)$ decreases. Thus, as $i$
increases, $R(g_i)$ decreases. We obtain that $2z\cdot R(g_h)\geq
|f(i)|\geq 2z\cdot R(g_{t-1})$.

Now consider the solution $D_c(j+1)$ obtained by doing reverse
operations on $D_c(j)$ with $s_{j+1}$ and $G_j$. Without of loss of
generality, suppose $o(s_{j+1})$ is used to cover gaps from $g_t$ to
$g_{t'}$ in $G_j$. With the similar analysis as above, we can obtain
$|f(j+1)|\leq 2z\cdot R(g_t)$, regardless of whether
the right endpoint of $I(s_{j+1})$ is at $\beta$ in $D_c(j+1)$.
Note that $R(g_t)\leq R(g_{t-1})$.
To see this, on one hand, it holds that $C(o_t)\leq C(o_{t-1})$. On the other hand,
since $g_t$ is to the left of $g_{t-1}$ and $s_{j+1}$ is to the right
of $s_j$, the number of sensors between
$g_t$ and $s_{j+1}$ is larger than that of the sensors between
$g_{t-1}$ and $s_j$, i.e., $C(s_{j+1},g_{t}) > C(s_{j},g_{t-1})$.
Hence $R(g_t)\leq R(g_{t-1})$ holds.

The above discussion leads to
$|f(i)|\geq 2z\cdot R(g_{t-1})\geq 2z\cdot R(g_t)\geq |f(j+1)|$.

The claim thus follows.
\end{myproof}

\section{The General Case}
\label{sec:general}
In this section, we consider the general case where sensors may be
everywhere on $L$. We present an $O(n\log n)$ time algorithm by
generalizing our algorithmic techniques for the one-sided case.

We assume there is at least one sensor whose covering interval
intersects $B$. The case where this assumption does not hold can be solved using similar but
simpler techniques and we will handle this case at the end of this section in Lemma \ref{lem:180}.

Let $s_l$ (resp., $s_r$) be the leftmost (resp.,
rightmost) sensor whose covering interval intersects $B$. We assume
$1<l$ and $r<n$, since otherwise it becomes the one-sided case. Let
$S_L=S(1,l-1)$,  $S_I=S(l,r)$, and $S_R=S(r+1,n)$.

We first give some intuition on how the problem can be solved.
Suppose $D_{opt}$ is an optimal solution. If no sensors of $S_L$ have been moved in $D_{opt}$, then we can compute $D_{opt}$ by solving a one-sided case on the sensors in $S(l,n)$. Similarly, if no sensors of $S_R$ have been moved in $D_{opt}$, then we can compute $D_{opt}$ by solving a one-sided case on the sensors in $S(1,r)$. Otherwise, there are sensors in both $S_L$ and $S_R$ that have been moved in $D_{opt}$. For this case, our main effort will be finding $l^*$, where $l^*$ is the smallest index such that sensor $s_{l^*}$ has been moved in $D_{opt}$. Note that $l^*\leq l-1$.
By the definition of $l^*$, sensors in $S(1,l^*-1)$ are useless for computing $D_{opt}$. Further, due to the order preserving property, the sc-intervals of sensors of $S(l^*,l-1)$ must all intersect $B$ in $D_{opt}$. Hence,
after we have $l^*$, $D_{opt}$ can be computed as follows.
We first ``manually'' move each sensor $s_i$ for $l^*\leq i\leq l-1$ rightwards to $-z$ and then apply our one-sided case algorithm on the sensors in $S(l^*,n)$, and the obtained solution is
$D_{opt}$.

\subsection{Observations}

Let $\lambda=\lceil\frac{\beta}{2z}\rceil$, i.e., the minimum number of
sensors necessary to fully cover $B$.

We introduce a few new definitions.
Consider any $i$ with $1\leq i\leq l$ and any $j$ with $r\leq j\leq n$ such
that $j-i+1\geq \lambda$. If $i\neq l$, define
$D^L_s(i,j)=\sum_{t=i}^{l-1}(-z-x_t)$, i.e., the total sum of the
moving distances for ``manually'' moving all sensors in $S(i,l-1)$
rightwards to $-z$ (such that the right endpoints of their covering
intervals are all at $0$); otherwise, $D^L_s(l,j)=0$.
Similarly, if $j\neq r$, define
$D^R_s(i,j)=\sum_{t=r+1}^j(x_t-z-\beta)$; otherwise, $D^R_s(i,r)=0$.
Let $D_s(i,j)=D_s^L(i,j)+D_s^R(i,j)$. Let $F(i,j)$ denote the
configuration after the above manual movements and including only sensors
in $S(i,j)$. Hence, $F(i,j)$ is an instance of the containing case on
sensors in $S(i,j)$.
Let $D_c(i,j)$ be the solution obtained by
applying our containing case algorithm on $F(i,j)$.
Finally, let $D(i,j)=D_c(i,j)+D_s(i,j)$.
For simplicity, for any $i$ and $j$ with $j-i+1<\lambda$, we let
$D(i,j)=+\infty$,  as $S(i,j)$ does not have enough sensors to
fully cover $B$.

For each $i$ with $1\leq i\leq l$, define $f(i)$ to be the index in
$[r,n]$ such that $D(i,f(i))=\min_{r\leq j\leq n}D(i,j)$.
Similarly, for each $j$ with $r\leq j\leq n$, define $f(j)$ to be the index in
$[1,l]$ such that $D(f(j),j)=\min_{1\leq i\leq l}D(i,j)$.

%

Let $D_{opt}$ denote the optimal solution. We have
the following lemma.

\begin{lemma}\label{lem:90}
	$D_{opt}=\min_{1\leq i\leq l, r\leq j\leq n}D(i,j)=\min_{1\leq
	i\leq l}D(i,f(i))=\min_{r\leq j\leq n}D(f(j),j)$.
\end{lemma}
\begin{myproof}
We assume at least one sensor in $S_L$ and at least one sensor in $S_R$ are
moved in $D_{opt}$, since other cases can be proved similarly (but
in a simpler way).

Let $l^*$ be the index of the leftmost sensor in $S_L$ that is moved in
$D_{opt}$, and let $r^*$ be index of the rightmost sensor in $S_R$ that is
moved in $D_{opt}$. Clearly, the covering intervals of $s_{l^*}$ and
$s_{r^*}$ must intersect $B$ in $D_{opt}$.  By the order preserving
property, all sensors in $S(l^*,l-1)$ are moved such that their
covering intervals in $D_{opt}$ all intersect $B$, and similarly,
all sensors in $S(r+1,r^*)$ are moved such that their
covering intervals in $D_{opt}$ all intersect $B$. Therefore, we can
obtain $D_{opt}$ by first manually moving sensors in $S(l^*,l-1)$
rightwards to $-z$ and moving sensors in $S(r+1,r^*)$ leftwards to
$\beta+z$, and then apply our containing case algorithm on
$S(l^*,r^*)$ (and the obtained solution is $D_{opt}$). According to our
definition of $D(i,j)$, we have $D_{opt}=D(l^*,r^*)$.

Therefore, it holds that $D_{opt}=\min_{1\leq i\leq l, r\leq j\leq
n}D(i,j)$. The definitions of $f(i)$ and $f(j)$ immediately lead to
$D_{opt}=\min_{1\leq i\leq l}D(i,f(i))=\min_{r\leq j\leq n}D(f(j),j)$.
The lemma thus follows.
\end{myproof}

Let $l^*$ and $r^*$ be the indices with $1\leq l^*\leq l$ and $r\leq
r^*\leq n$ such that $D(l^*,r^*)=D_{opt}$. It is easy to
see that $l^*=f(r^*)$  and $r^*=f(l^*)$.

To compute $D_{opt}$, if we know either $l^*$ or $r^*$, then $D_{opt}$
can be computed in additional $O(n\log n)$ time, as follows.
Suppose $l^*$ is known to us. We first
``manually'' move each sensor $s_i$ for $l^*\leq i\leq l-1$ rightwards
to $-z$ (this step is not necessary for the case $l^*=l$) and then apply
our one-sided case algorithm on $S(l^*,n)$ (the obtained solution is
$D_{opt}$).  Hence, the key is to determine $l^*$ or $r^*$.

\subsection{The Case $|S_I|\geq \lambda$}

First, we show that if $|S_I|\geq \lambda$,
then we can easily compute $l^*$ and $r^*$ in $O(n\log n)$ time by using the following lemma.

\begin{lemma}\label{lem:100}
If $|S_I|\geq \lambda$, then it holds that $f(i)=r^*$ for any $i\in
[1,l]$ and $f(j)=l^*$ for any $j\in [r,n]$.
\end{lemma}
\begin{myproof}
We only prove the former case since the latter case can be proved
similarly. Due to $|S_I|\geq \lambda$, we have $2z\cdot|S_I|\geq
\beta$.  Hence, we can run our containing case algorithm
on $S_I$ to obtain a solution that covers $B$ fully, which is $D_c(l,r)$
according to our definition.  Depending on whether $2z\cdot|S_I|= \beta$,
there are two cases. In the following, we first prove the case with
$2z\cdot|S_I|> \beta$.

If $2z\cdot |S_I|> \beta$, there must exist an overlap, denoted by
$o$, in the configuration of $D_c(l,r)$.
Note that $o$ may be a subset of an original overlap in the input. In
the following, we assume $o$ has two generators since the case where
$o$ has only one generator can be proved similarly but in a much
simpler way. Let $s_k$ and $s_{k+1}$ be the generators of the overlap
$o$.


To compute $f(l)$, we can apply our one-sided case algorithm on the
sensors $S(l,n)$. Recall the our one-sided case algorithm works by
doing the reverse processes on the configuration $D_c(l,r)$ and considering
sensors in $S_{R}$ one by one from left to right. Consider any $j$ with $r+1\leq j\leq n$.
According to the reverse operations, since $o$ is an
overlap in $D_c(l,r)$, comparing the two configurations $D_c(l,r)$ and
$D_c(l,j)$, sensors in $S(r+1,j)$ are used to cover some gaps of $D_c(l,r)$ that are
to the right of the overlap $o$.
Hence, for each sensor in $S(l,k)$, its locations in $D_c(l,r)$ and
$D_c(l,j)$ are the same, and in other words, $D_c(l,j)$ is determined
by the locations of the sensors of $S(k+1,r)$ in $D_c(l,r)$. This
implies that the index $f(i)$ is only determined by the locations of
the sensors of $S(k+1,r)$ in $D_c(l,r)$.

Consider the configuration $D_c(l-1,r)$. Comparing with $D_c(l,r)$, we
have one more sensor $s_{l-1}$ on the left side of $B$.
Hence, $s_{k}$ and $s_{k+1}$
still define an overlap in $D_c(l-1,r)$:
Although $s_{k}$ in $D_c(l-1,r)$ may be strictly to the right of its location in
$D_c(l,r)$ (in this case, the new overlap is longer than $o$), the
sensor $s_{k+1}$ has the same position in $D_c(l,r)$ and $D_c(l-1,r)$.
This also implies that
each sensor of $S(k+1,r)$ has the same position in $D_c(l,r)$ and $D_c(l-1,r)$.

We have shown that the index $f(l)$ is only determined by the
locations of the sensors of $S(k+1,r)$. Since each
sensor of $S(k+1,r)$ has the same location in $D_c(l,r)$ and $D_c(l-1,r)$,
and $s_k$ and $s_{k+1}$ define an overlap in both configurations,
if we do reverse operations on $D_c(l-1,r)$ and $S_R$ to compute
$f(l-1)$, we will obtain the same result as that for $D_c(l,r)$ and
$S_R$, i.e., $f(l-1)=f(l)$.

By similar analysis, we can show that
$f(l)=f(l-1)=\cdots=f(1)$, which leads to the lemma for the case where
$2z\cdot|S_I|>\beta$.

In the following, we prove the case with $2z\cdot|S_I|=\beta$. The
proof is similar in spirit to the first case.

In this case, all sensors in the configuration of $D_c(l,r)$ has to be in
attached positions.
If sensors in $S(l,r)$ do not define any overlap in the
input configuration, then these sensors must be in attached positions
and exactly cover $B$, implying that $D_{opt}=0$ and
$f(i)=r^*$ for each $1\leq i\leq l$, and thus the lemma follows.
Otherwise, suppose we apply our containing case algorithm on $S(l,r)$
to compute $D_c(l,r)$ and let $o$ be the overlap used to cover a gap
$g$ in the last shift process of the algorithm. Note that $|o|=|g|$ due to $2z\cdot|S_I|=\beta$.


We assume $o$ has two generators since the case where $o$ has only one
generator can be proved similarly (but in a simpler way). Let $s_k$
and $s_{k+1}$ be the left and right generators of $o$, respectively.
Another way to think of $o$ is that if the length of $B$ was
$\beta-\epsilon$ for an infinitesimal value $\epsilon$, then there would
be an overlap defined by $s_k$ and $s_{k+1}$ in $D_c(l,r)$.

According to the one-sided case algorithm, we can obtain $f(l)$ by doing reverse
operations on $D_c(l,r)$ and considering the sensors in $S_R$ from left to right.
One observation is that each sensor $s_i$ in $S(l,k)$ has
the same location in $D_c(l,r)$ and $D_c(l,f(l))$.
To see this, according to our reverse operations, if sensors in
$S(r+1,f(l))$ are used to cover some gaps of $D_c(l,r)$, then these
gaps must be to the right of $o$ since $o$ is used to cover the last
gap $g$, and after these reverse operations, some sensors to the right
of $o$ may have been moved leftwards but no sensor to the left of $o$ is
moved. In other words, the index $f(l)$ is determined only by the
locations of sensors of $S(k+1,r)$ in the configuration $D_c(l,r)$.

Now consider the solution $D_c(l-1,r)$. Similarly to the
analysis in the first case, for each sensor in $S(l,k)$, its
location in $D_c(l-1,r)$ may be strictly to the right of its location
in $D_c(l,r)$. However, each sensor in $S(k+1,r)$ has the same location in
$D_c(l,r)$ and $D_c(l-1,r)$. Therefore, if we do reverse
operations on $D_c(l-1,r)$ and $S_R$ to compute $f(l-1)$, we will
obtain the same result as that for $D_c(l,r)$ and $S_R$, i.e.,
$f(l-1)=f(l)$. Similar analysis can prove that $f(l)=f(l-1)=\cdots=f(1)$.

The lemma thus follows.
\end{myproof}

By Lemma \ref{lem:100}, if $|S_I|\geq \lambda$, then
it holds that $f(1)=r^*$, which can be easily computed
in $O(n\log n)$ time by applying our one-sided case algorithm on
$S(1,n)$ after moving sensors in $S_L$ rightwards to the position $-z$.

In the following discussion, we assume $|S_I|<\lambda$.
Note that $|S(l^*,r^*)|\geq \lambda$ always holds.
Since both $|S(l^*,r^*)|$ and $\lambda$ are
integers, either $|S(l^*,r^*)|\geq \lambda+1$ or
$|S(l^*,r^*)|= \lambda$. We have different algorithms for these
two subcases.

\subsection{The Case $|S_I|<\lambda$ and $|S(l^*,r^*)|\geq \lambda+1$}

The subcase $|S(l^*,r^*)|\geq\lambda+1$ can be easily handled due to
the following lemma, which is proved based on the unimodal property described in Lemma \ref{lem:80}.

\begin{lemma}\label{lem:110}
If $|S(l^*,r^*)|\geq \lambda+1$, then $f(i)=r^*$ holds for any $i$ with $1\leq i<l^*$.
\end{lemma}
\begin{myproof}
We assume $1<l^*<l$ and $r<r^*<n$ since other cases can be proved using
similar techniques but in simpler ways. Recall that $r^*=f(l^*)$.

Since $|S(l^*,r^*)|\geq \lambda+1$, $|S(l^*,r^*-1)|\geq \lambda$ holds, and thus,
$D_c(l^*,r^*-1)\neq +\infty$. We can obtain the
solution $D_c(l^*,r^*)$ by doing reverse operations on
$D_c(l^*,r^*-1)$ with sensor $s_{r^*}$. Let
$R(l^*,s_{r^*})=D_c(l^*,r^*-1)-D_c(l^*,r^*)$, and as in Section
\ref{sec:onesided}, we can consider $R(l^*,s_{r^*})$ as the revenue or savings incurred by $s_{r^*}$ on $D_c(l^*,r^*-1)$.
For any sensor
$s_j\in S_R$, let $d(s_j)=x_j-z-\beta$. By the definition of
$f(l^*)$ ($=r^*$), if we consider finding an optimal solution for the one-sided
case on $S(l^*,r^*-1)$ after sensors in $S(l^*,l-1)$ are moved to $-z$
and sensors in $S(r+1,r^*-1)$ are moved to $\beta+z$, then we have
$R(l^*,s_{r^*})\geq d(s_{r^*})$.

Similarly, we can also obtain $D_c(l^*,r^*+1)$ by doing reverse operations on
$D_c(l^*,r^*)$ with sensor $s_{r^*+1}$, and let
$R(l^*,s_{r^*+1})=D_c(l^*,r^*)-D_c(l^*,r^*+1)$. Again, by the
definition of $f(l^*)$  ($=r^*$), it holds that $R(l^*,s_{r^*+1})\leq d(s_{r^*+1})$.

In the following, we first prove $f(l^*-1)=r^*$.

Similarly as above, define
$R(l^*-1,s_{r^*})=D_c(l^*-1,r^*-1)-D_c(l^*-1,r^*)$ and
$R(l^*-1,s_{r^*+1})=D_c(l^*-1,r^*)-D_c(l^*-1,r^*+1)$. By the unimodal
property in Lemma \ref{lem:80}, to prove $f(l^*-1)=r^*$, it is sufficient to show that
$D(l^*-1,s_{r^*})-D(l^*-1,s_{r^*-1})\leq 0$ and
$D(l^*-1,s_{r^*})-D(l^*-1,s_{r^*+1})\leq 0$. Note that
$D(l^*-1,s_{r^*})-D(l^*-1,s_{r^*-1})=d(s_{r^*})-R(l^*-1,s_{r^*})$ and
$D(l^*-1,s_{r^*})-D(l^*-1,s_{r^*+1})=R(l^*-1,s_{r^*+1})-d(s_{r^*+1})$.
Therefore, to prove $f(l^*-1)=r^*$, it suffices to show that
$R(l^*-1,s_{r^*})\geq d(s_{r^*})$ and $R(l^*-1,s_{r^*+1})\leq d(s_{r^*+1})$.
To this end, in the sequel we show that $R(l^*-1,s_{r^*})=R(l^*,s_{r^*})$
and $R(l^*-1,s_{r^*})=R(l^*,s_{r^*})$, which will lead to the lemma.

The proof techniques are similar to those used in the proof of Lemma
\ref{lem:100}. We first prove $R(l^*-1,s_{r^*})=R(l^*,s_{r^*})$.
Since $|S(l^*,r^*)| = r^*-l^*+1\geq
\lambda+1$, it holds that $2z(r^*-l^*)\geq \beta$.  As in the proof of
Lemma \ref{lem:100}, depending on whether $2z(r^*-l^*)>\beta$ or
$2z(r^*-l^*)=\beta$, there are two cases.

\begin{enumerate}
\item
If $2z(r^*-l^*)>\beta$, then the configuration $D_c(l^*,r^*-1)$
must have an overlap $o$. We assume $o$ has two generators $s_k$ and
$s_{k+1}$ since the other case where it has only one generator
can be proved similarly but in a simpler way.
Consider the two configurations $D_c(l^*,r^*-1)$ and
$D_c(l^*,r^*)$. We can obtain $D_c(l^*,r^*)$ by doing reverse operations on
$D_c(l^*,r^*-1)$ and sensor $s_{r^*}$.

As in the proof of Lemma \ref{lem:100}, due to the overlap $o$, each sensor of $S(l^*,k)$ has the same location in
$D_c(l^*,r^*-1)$ and
$D_c(l^*,r^*)$. Hence, the value $R(l^*,s_{r^*})$
only depends on the locations of the sensors of $S(k+1,r^*-1)$ in
$D_c(l^*,r^*-1)$.

As in Lemma \ref{lem:100}, sensors $s_k$ and $s_{k+1}$ still define an overlap in $D_c(l^*-1,r^*-1)$. Hence, each sensor of $S(k+1,r^*-1)$ has the same location in $D_c(l^*-1,r^*-1)$ and $D_c(l^*,r^*-1)$. Similarly, the value $R(l^*-1,s_{r^*})$
only depends on the locations of the sensors of $S(k+1,r^*-1)$ in
$D_c(l^*-1,r^*-1)$.

Therefore, $R(l^*-1,s_{r^*})=R(l^*,s_{r^*})$ holds.

\item
If $2z(r^*-l^*)=\beta$, then in the configuration $D_c(l^*,r^*-1)$ all
sensors of $S(l^*,r^*-1)$ are in attached position. Suppose we compute $D_c(l^*,r^*-1)$ by
using our containing case algorithm; as in Lemma \ref{lem:100}, let $o$ be
the overlap used to cover a gap $g$ in the last shift process of the algorithm.
Again, we assume $o$ has two generators $s_k$ and $s_{k+1}$.

As the analysis in Lemma \ref{lem:100} and the above case, the
value $R(l^*,s_{r^*})$
only depends on the locations of the sensors of $S(k+1,r^*-1)$ in
$D_c(l^*,r^*-1)$ and $R(l^*-1,s_{r^*})$
only depends on the locations of the sensors of $S(k+1,r^*-1)$ in
$D_c(l^*-1,r^*-1)$. Further, each sensor of $S(k+1,r^*-1)$ has the same location in $D_c(l^*-1,r^*-1)$ and $D_c(l^*,r^*-1)$.
Thus, $R(l^*-1,s_{r^*})=R(l^*,s_{r^*})$ holds.
\end{enumerate}

The above proves that $R(l^*-1,s_{r^*})=R(l^*,s_{r^*})$.

To prove $R(l^*-1,s_{r^*+1})=R(l^*,s_{r^*+1})$, we can use the similar
techniques. Note that since $|S(l^*,r^*)|\geq \lambda+1$, we
have $2z\cdot (r^*-l^*+1) >\beta$, and thus we only need to consider
the above first case. We omit the details.

The lemma is thus proved.
\end{myproof}

By Lemma \ref{lem:110}, if $|S(l^*,r^*)|\geq \lambda+1$, then
it holds that $f(1)=r^*$, which can be easily computed
in $O(n\log n)$ time by applying our one-sided case algorithm on
$S(1,n)$ after moving sensors in $S_L$ rightwards to the position $-z$.

\subsection{The Case $|S_I|<\lambda$ and $|S(l^*,r^*)|=\lambda$}
\label{subsec:equallambda}

It remains to handle the case where $|S(l^*,r^*)|=\lambda$.
Due to $l^*\leq l$ and $r^*\geq r$, we have
$\max\{1,r-\lambda+1\}\leq l^*\leq \min\{l,n-\lambda+1\}$.
In the following,
for simplicity of discussion, we assume $r-\lambda+1>1$ and $l<n-\lambda+1$
since the other cases can be solved similarly. Let $l'=r-\lambda+1$.
Thus, we have $l'\leq l^*\leq l$, and for any $i$ with $i\geq 0$ and $r+i\leq n$, $|S(l'+i,r+i)|=\lambda$.
Clearly, $D_{opt}=\min_{0\leq i\leq l-l'}D(l'+i,r+i)$. Let $l''=l-l'$.

In the following, we present an $O(n\log n)$ time algorithm that can
compute the solutions $D(l'+i,r+i)$ for all $i=0,1,\ldots,l''$.
Recall that $D(l'+i,r+i)=D_c(l'+i,r+i)+D_s(l'+i,r+i)$.
We can easily compute  $D_s(l'+i,r+i)$ for
all $i=0,1,\ldots,l''$ in $O(n)$ time. Therefore, it is sufficient to
compute the solutions $D_c(l'+i,r+i)$ for all $i=0,1,\ldots,l''$ in
$O(n\log n)$ time, which is our focus below. To simplify the notation,
we use $D_c(i)$ to represent $D_c(l'+i,r+i)$.

In the following discussion, unless otherwise stated, we assume
all sensors in $S(1,l-1)$ are at $-z$ and all sensors in
$S(r+1,n)$ are at $\beta+z$; sensors in $S(l,r)$ are in their original locations as input. In other words, we work on the configuration $F(1,n)$.

\subsubsection{The case $\lambda=\frac{\beta}{2z}$}

We first consider a special case where $\lambda=\frac{\beta}{2z}$,
i.e., $\frac{\beta}{2z}$ is an integer. In this case, for each $0\leq i\leq l''$,
the configuration
$D_c(i)$ has a very special pattern: sensors in $S(l'+i,r+i)$ are in
attached positions with $s_{l'+i}$ at $z$.
The following lemma gives an $O(n\log n)$ time algorithm
for this special case.

\begin{lemma}\label{lem:120}
If $|S(l^*,r^*)|=\lambda$ and $\lambda=\frac{\beta}{2z}$,
	we can compute $D_{opt}$ in $O(n\log n)$ time.
\end{lemma}
\begin{myproof}
For any configuration $F$, we define its {\em aggregate-distance}
as the sum of the distances of all sensors between their locations in $F$
and their locations in $F(1,n)$. Note that in $F(1,n)$, sensors of $S(1,l-1)$ have been moved to $-z$ and sensors of $S(r+1,n)$ have been moved to $\beta+z$.

We first compute $D_c(0)$, i.e., $D_c(l',r)$, which can be done in $O(n)$ time.
$D_c(1)$ can be obtained from the configuration $D_c(0)$
by moving each sensor in $S(l',r+1)$ leftwards by distance
$2z$. In general, for each $l'\leq i\leq l''$, we can obtain
the configuration $D_c(i+1)$ from $D_c(i)$ by moving each sensor in $S(l'+i,r+i+1)$
leftwards by distance $2z$. To compute the value  $D_c(i+1)$
efficiently, however, we need to do the above
movement carefully, as follows.

Let $S_n$ be the set of sensors of $S(l',r)$ whose displacements in the configuration
$D_c(0)$ are negative with respect to their
locations in $F(1,n)$ (i.e., their locations in $D_c(0)$ are strictly to the right of their locations in $F(1,n)$),
and let $k_n=|S_n|$. Since the displacement of $s_{r+1}$ is not negative, the number of
sensors of $S(l',r+1)$  with non-negative
displacements in $D_c(0)$ is $\lambda+1-k_n$. If we move all
sensors of $S(l',r+1)$ leftwards by an infinitesimal distance $\delta$ such that
the displacement of each sensor in $S_n$ is still negative after the
movement, then the aggregate-distance of the new configuration is
$D_c(0)+\delta\cdot (\lambda+1-k_n-2k_n)$. If we keep moving, then the
displacements of some sensors in $S_n$ will become zero, at which
moments we should update the value $k_n$ for later computation. We stop the algorithm after $\delta$ becomes $2z$, at which moment $D_c(1)$ is obtained.

We can use the similar idea to obtain $D_c(2)$ and so on until
$D_c(l'')$.
By careful implementation, we can compute all these solutions in
$O(n\log n)$ time as follows.

First, we compute $D_c(0)$ and obtain the set $S_n$ and $k_n$.
Let $A$ be the set of the absolute values of the
displacements of sensors in $S_n$.
Furthermore, let $A=A\cup \{2z\cdot i\ | \ 1\leq i\leq l''\}$.
For simplicity of discussion, we assume no two
values in $A$ are the same.

We sort the values in $A$ in increasing order. Starting from the
configuration $D_c(0)$,
our algorithm ``sweeps'' a value 
$\delta$ from zero to $2z\cdot l''$ and $\delta$ represents
the total leftwards movement made so far by our algorithm.
Note that after moving the distance of $2z\cdot l''$, we will obtain the
configuration $D_c(l'')$.

During the algorithm, when
$\delta$ is equal to any value in $A$, an event happens and we need to
update the value $k_n$ accordingly. In general, suppose we have
computed the aggregate-distance $M(\delta_1)$ of the current configuration at distance
$\delta=\delta_1$ and we also
know the current value $k_n(\delta_1)$. Initially, $M(0)=D_c(0)$ and $k_n(0)$
is known. Consider the next event $\delta=\delta_2$. First, we
compute the aggregate-distance $M(\delta_2)=M(\delta_1)+(\delta_2-\delta_1)\cdot
(\lambda+1-3k_n(\delta_1))$. If $\delta_2$ is
equal to the absolute displacement of a sensor in $S_n$, then we
update $k_n(\delta_2)=k_n(\delta_1)-1$. Otherwise, $\delta_2=2z\cdot i$
for some $1\leq i\leq l''$, and in this case, we obtain $D_c(i)=M(\delta_2)$ and
 $k_n(\delta_2)=k_n(\delta_1)$.

In this way, each event takes $O(1)$ time. There are $O(n)$ events.
Hence, we can compute $D_c(i)$ for $i=0,1,\ldots,l''$ in $O(n\log n)$ time.

Since we already have the values $D_s(i)$ for $i=0,1,\ldots,l''$, we can obtain the values $D(i)$ for all $i=0,1,\ldots,l''$ and $D_{opt}$ in additional $O(n)$ time.
\end{myproof}

\subsubsection{The case $\lambda\neq \frac{\beta}{2z}$}

In the following, we assume $\lambda\neq \frac{\beta}{2z}$, i.e.,
$\frac{\beta}{2z}$ is not an integer. This implies
that there must be an overlap in any solution $D_c(i)$ for $0\leq
i\leq l''$.

We first use our containing case algorithm to compute $D_c(0)$ on the configuration $F(l',r)$ with sensors in $S(l',r)$.
Below, we present an algorithm that can compute $D_c(1)$ by modifying the configuration
$D_c(0)$. The algorithm consists of two main steps.



The first main step is to compute $D_c(l',r+1)$ by doing reverse operations on
$D_c(l')$ with sensor $s_{r+1}$ at $\beta+z$. This is done in the same way as in our
 one-sided case algorithm.

The second main step is to compute $D_c(1)$ by modifying the configuration $D_c(l',r+1)$, as follows.

Note that $D_c(1)$ is on the configuration $F(l'+1,r+1)$ with sensors
in $S(l'+1,r+1)$ while $D_c(l',r+1)$ is on $F(l',r+1)$ with sensors
in $S(l',r+1)$. Hence, $s_{l'}$ is not used in $D_c(1)$ but
may be used in $D_c(l',r+1)$. If in $D_c(l',r+1)$, $s_{l'}$ covers
some portion of $B$ that is not covered by any other sensor in
$S(l'+1,r+1)$, then we should move sensors of $S(l'+1,r'+1)$ to cover
the above portion and more specifically, that portion should be
covered by eliminating some overlaps in $D_c(l',r+1)$. The details are
given below.

Consider the configuration $D_c(l',r+1)$.
If $s_{l'}$ is at $-z$, then $I(s_{l'})\cap B=\emptyset$ and $B$ is
covered by sensors of $S(l'+1,r+1)$, implying that $D_c(1)=D_c(l',r+1)$.

If $s_{l'}$ is not at $-z$, then let $g=I(s_{l'})\cap B$. The following lemma implies that $s_{l'}$ is the only sensor that covers $g$ in $D_c(l',r+1)$.

\begin{lemma}\label{lem:130}
No sensor in $S(l'+1,r'+1)$ covers $g$ in $D_c(l',r+1)$.
\end{lemma}
\begin{myproof}
Recall that all sensors in $S(l',l-1)$ are initially at $-z$.
Since $|S(l',r)|=\lambda$, $s_{l'}$ must be strictly to the right of $-z$ in $D_c(l')$ (i.e., $s_{l'}$ has been moved rightwards). Due to the order preserving property,
sensors in $S(l',l)$ must be in attached positions in $D_c(l')$. When we compute $D_c(l',r+1)$, sensor
$s_{l'}$ may be moved leftwards due to the reverse operations, in which
case all sensors in $S(l',l)$ must be moved leftwards by the same
amount because they were in attached positions in $D_c(l')$. Hence,
sensors in $S(l',l)$ are also in attached positions in $D_c(l',r+1)$,
which implies that $g$ is only covered by $s_{l'}$ in $D_c(l',r+1)$.
\end{myproof}

To obtain $D_c(l'+1)$, we first remove $s_{l'}$ and then
cover $g$ by eliminating overlaps of $D_c(l',r+1)$ from left to right until $g$ is fully covered. Specifically, let $o_1,o_2,\ldots,o_k$ be the overlaps of $D_c(l',r+1)$ sorted from left to right. We move the sensors between $g$ and $o_1$ leftwards by
distance $\min\{|g|,|o_1|\}$.
This movement can be done in $O(\log n)$ time by updating the position tree $T_p$. If $|g|\leq |o_1|$, then we are done.
Otherwise, we consider the next overlap $o_2$. We continue this
procedure until $g$ is fully covered. Note that since
$|S(l'+1,r+1)|=\lambda$, it holds that $\sum_{i=1}^k|o_i|\geq |g|$,
implying that $g$ will eventually be fully covered. Let $D$ be the obtained
configuration. The following lemma shows that $D$ is $D_c(1)$.

\begin{lemma}\label{lem:140}
$D$ is $D_c(1)$.
\end{lemma}
\begin{myproof}
Suppose we run our containing case algorithm on both configurations
$F(l'+1,r+1)$ and $F(l',r+1)$ simultaneously by considering the sensors
in the two sets in the order {\em from right to left}, to compute $D_c(1)$ and $D_c(l',r+1)$, respectively. Let the algorithm on $F(l'+1,r+1)$ be $A_{l'+1}$ and let the algorithm on $F(l',r+1)$ be $A_{l'}$.

Let $g_1$ be the first gap such that $A_{l'}$ and $A_{l'+1}$ use different overlaps to cover it. Let $F$ be the configuration in $A_{l'}$ including only sensors in $S(l'+1,r+1)$ right before $g_1$ is considered. Hence, $A_{l'+1}$ has the same configuration as $F$ right before $g_1$ is considered. Below, if the context is clear, we use $F$ to refer to the configurations in both  $A_{l'}$  and $A_{l'+1}$.

Since the only difference of $F(l'+1,r+1)$ and $F(l',r+1)$ is that
$F(l',r+1)$ has an additional overlap $o$ of size $2z$ defined by
$s_{l'}$ at $-z$, $g_1$ must be covered by $o$ in $A_{l'}$ while $g_1$ is
covered by other overlaps in $A_{l'+1}$. Since $o$ is the leftmost overlap in $F(l',r+1)$, there are no overlaps between $o$ and $g_1$ in $F$. Since both algorithms consider the gaps from right to left, the remaining gaps in
$F$ are all to the left of $g_1$, and let $G$ denote the set of
the remaining gaps in $F$ and let $d_G$ denote the total sum of the lengths
of all these gaps.

%

In algorithm $A_{l'+1}$, all overlaps
are to the right of $g_1$ in $F$, and thus, according to our
containing case algorithm, these overlaps will be used
from left to right to cover the gaps of $G$ until all gaps are
covered and the total sum of the overlaps eliminated is exactly $d_G$.
The obtained solution is $D_c(l'+1)$.

In algorithm $A_{l'}$, however, depending on the costs, we can use either the
overlaps to the right of $g_1$ or use $o$ to cover the gaps of $G$.
Consider the configuration $F(l',r+1)$.  Recall that in $F(l',r+1)$,
$I(s_{l'})$ covers a portion of $B$, denoted by $g$, which is not
covered by any sensor of $S(l'+1,r+1)$. This means that
in algorithm $A_{l'}$ the overlap $o$ eventually covers
some gaps of $G$ of total length $|g|$
and the rest of the gaps of $G$, whose total length is $d_G-|g|$, are
covered by the overlaps to the right $g_1$ in the order from left to right.

Now consider our algorithm for computing $D$ based on $D_c(l',r+1)$.
The overlaps of $D_c(l',r+1)$ are to the right of $g$, and we
obtain $D$ by eliminating these overlaps from
left to right until $g$ is fully covered (thus the total length of the
overlaps eliminated is $|g|$).

Combining the discussion of the last two paragraphs, it is equivalent to
say that $D$ is obtained from $F$ by covering the gaps of $G$
by eliminating the overlaps of $F$
from left to right with a total length of $d_G-|g|+|g|=d_G$.
Therefore, the configuration $D$ is exactly the same as the
configuration $D_c(l'+1)$.

The lemma thus follows.
\end{myproof}

The above gives a way to compute $D_c(1)$ from $D_c(0)$. In
general,  for each $0\leq i\leq l''$, if we know $D_c(i)$,
we can use the same approach to compute $D_c(i+1)$ and the proof of the correctness is similar as in Lemma \ref{lem:140}.

We say a solution $D_c(i)$ for $i\in [0,l'']$ is {\em trivial} if the
coordinate of the
right endpoint of $I(s_{r+i})$ is strictly larger than $\beta$.
By using the similar algorithm as in Lemma \ref{lem:120}, we have the following lemma.

\begin{lemma}\label{lem:150}
Suppose $k$ is the smallest index in $[0,l'']$ such that $D_c(k)$ is a trivial solution; then we can compute $D_c(i)$ for all $i=k,k+1,\ldots, l''$ in $O(n\log n)$ time.
\end{lemma}
\begin{myproof}
Let $k$ be the index specified in the lemma statement. Let $x$ be the
coordinate of the right endpoint of $I(s_{r+i})$ in the configuration
$D_c(k)$. Since $x>\beta$,
sensor $s_{r+k}$ defines an overlap $[\beta,x]$ in $D_c(k)$.

We can obtain $D_c(l'+k,r+k+1)$ by doing the
reverse operations on $D_c(k)$ and sensor $s_{r+k+1}$. Since
$s_{r+k}$ already defines an overlap $[\beta,x]$ that is to the right of $\beta$, the
configuration $D_c(l'+k,r+k+1)$ is exactly the same as $D_c(k)$ except that
$D_c(l'+k,r+k+1)$ includes $[\beta,\beta+2z]$ as an overlap defined by
sensor $s_{r+k+1}$.

As the way we compute $D_c(1)$ from $D_c(l',r+1)$, we can compute $D_c(k+1)$ by modifying the configuration $D_c(l'+k,r+k+1)$ in the
following way.
Let $g=I(s_{l'+k})\cap B$. To obtain $D_c(k+1)$, we remove $s_{l'+k}$ and cover $g$ by eliminating the overlaps of $D_c(l'+k,r+k+1)$ from left to right until
$g$ is covered.

The above computes $D_c(k+1)$ from $D_c(l'+k,r+k+1)$ .
Next, we show that
$D_c(k+1)$ has a very special pattern:
sensors of $S(l'+k+1,r+k+1)$ are in attached positions and
sensor $s_{l'+k+1}$ is at $z$ (i.e., the left endpoint of
$I(s_{l'+k+1})$ is at $0$).

Indeed, let $d_o$ be the sum of the lengths of all overlaps in $D_c(k)$.
Note that $|S(l'+k,r+k)|=\lambda$.
Since $2z\cdot (\lambda-1)<\beta$, sensors in $S(l'+k+1,r+k)$ are not
enough to fully cover $B$, which implies that $|g|>d_o$.
Recall that the two configurations $D_c(k)$ and $D_c(l'+k,r+k+1)$ are
the same except that the latter one has an additional overlap
$[\beta,\beta+2z]$. Consider the procedure for covering $g$
by eliminating the overlaps of $D_c(l'+k,r+k+1)$ from left to right.
Since $|g|>d_o$, all overlaps of $D_c(l'+k,r+k+1)$ except that
last one $[\beta,\beta+2z]$ will be eliminated, and the moment right
before the overlap $[\beta,\beta+2z]$ is used, sensors in
$S(l'+k+1,r+k+1)$ must be in attached positions. Finally, $D_c(k+1)$
is obtained after sensors of
$S(l'+k+1,r+k+1)$ are moved leftwards to cover $g$ completely,
which implies that all sensors of $S(l'+k+1,r+k+1)$ are in attached
positions in $D_c(k+1)$ and $s_{l'+k+1}$ is at $z$.
Further, since $2z\cdot \lambda>\beta$, the right endpoint of
$I(s_{r+k+1})$ is strictly to the right of $\beta$, implying that
$D_c(k+1)$ is a trivial solution.

Since $D_c(k+1)$ is also a trivial solution, by using
the similar analysis, we can show that for each $k+2\leq i\leq
l''$, $D_c(i)$ is a trivial solution and
has the following pattern:  sensors in $S(l'+i,r+i)$
are in attached positions with $s_{l'+i}$ at $z$.

Therefore, after $D_c(k+1)$ is computed, we can obtain all solutions
$D_c(i)$ for $k+2\leq i\leq l''$ by moving sensors leftwards.
We can use a similar sweeping algorithm as in
Lemma \ref{lem:120} to compute all these solutions $D_c(i)$ for $k+2\leq i\leq r$ in
$O(n\log n)$ time (we omit the details).

The lemma thus follows.
\end{myproof}

In the following, we compute solutions $D_c(i)$ for all
$i=0,1,\ldots, l''$ in $O(n\log n)$ time. Our algorithm will compute the
solutions $D_c(i)$ in the order from $0$ to $l''$ until either $D_c(l'')$ is obtained,
or we find a trivial solution and then we apply the algorithm in
Lemma \ref{lem:150}.

First, we compute $D_c(0)$ in $O(n\log n)$ time by applying our containing case algorithm
on the configuration $F(l',r)$. As in our one-sided case algorithm,
we also maintain the process information of the right-shift processes after
the last left-shift process in the above algorithm. Let $P=\{p_1,p_2,\ldots,p_q\}$ be the above
process list in the inverse time order (i.e., $p_1$ is the last
process of the algorithm), where $q$ is the number of these
processes. Let $G=\{g_1,g_2,\ldots,g_q\}$
and $O=\{o_1,o_2,\ldots,o_q\}$ be the corresponding gap list and
overlap list, i.e., for each $1\leq i\leq q$,
process $p_i$ covers $g_i$ by eliminating $o_i$. For each $1\leq i\leq
q$, we also maintain the cost $C(o_i)$ of the overlap $o_i$.
As discussed in Section \ref{sec:onesided}, the gaps of $G$ are sorted from right to left while the
overlaps of $O$ are sorted from left to right.
In addition, we maintain an extra overlap list
$O'=\{o_1',o_2',\ldots,o_h'\}$, which are the overlaps in the
configuration $D_c(0)$ sorted from left to right. The list $O'$ will
be used in the second main step for computing each $D_c(i)$.
According to their definitions, all overlaps of $O'$ are to the left of the overlaps of $O$.
As in the one-sided case algorithm, we only need
to use the position tree $T_p$ in the following algorithm.

To compute $D_c(1)$, the first main step is to compute $D_c(l',r+1)$ by doing the
reverse operations on $D_c(0)$ with $s_{r+1}$.
This step is the same as that in the one-sided case algorithm.
Let $o(s_{r+1})$ be the overlap $[\beta,\beta+2z]$ defined by
$s_{r+1}$ at $\beta+z$.
In general, suppose during the reverse operations
$g_1,g_2,\ldots,g_{t-1}$ are the gaps fully covered by $o(s_{r+1})$
and $g_t$ is only partially covered by a length of $d_t$. Then, gaps
$g_1,g_2,\ldots,g_{t-1}$ are removed from $G$, and $g_t$ is still in
$G$ but its length is changed to its original length minus $d_t$.
Correspondingly, the overlaps $o_1,o_2,\ldots,o_{t-1}$ are restored and $o_t$
is partially
restored with length $d_t$ in $D_c(l',r+1)$. We append $o_1,o_2,\ldots,o_{t}$ at the end
of $O'$. Since overlaps of $O'$ are to the left of overlaps of
$O$ and overlaps of the two lists $O$ and $O'$ are both sorted from
left to right, after the above ``append'' operation, the overlaps of the
new list $O'$ are still sorted from left to right.

The second main step is to compute $D_c(1)$ from $D_c(l',r+1)$,
by eliminating overlaps of $O'$ from left to right until
$I(s_{l'})\cap B$ is covered, as discussed earlier. For each overlap
that is eliminated, we remove it from $O'$, which can be done in
constant time.
Note that eliminating an overlap is essentially to move a subset of
consecutive sensors leftwards by the same distance, which takes
$O(\log n)$ time to update the position tree $T_p$.
Hence, the running time for this step is
$O((t'+1)\log n)$, where $t'$ is the number of overlaps that are eliminated
and the additional one is for the case where an overlap is not
completely eliminated while $I(s_{l'})\cap B$ is fully covered (at
which moment we obtain $D_c(1)$).

If $D_c(1)$ is a trivial solution, we are done. Otherwise, we
continue to compute $D_c(2)$, again by first computing
$D_c(l'+1,r+2)$ and then computing $D_c(2)$.
Let $G_1$ be the remaining gap list of $G$ after $D_c(1)$ is computed.
To compute $D_c(l'+1,r+2)$, we use $G_1$ to do the
reverse operations on $D_c(1)$ with $s_{r+2}$. Although
$G_1$ may not be the corresponding gap list for $D_c(1)$, Lemma \ref{lem:160} shows that the obtained result
using $G_1$ is $D_c(2)$, and further, this can be generalized to
$D_c(3), D_c(4),\ldots$ until $D_c(l'')$.

\begin{lemma}\label{lem:160}
Suppose $D_c(1)$ is not a trivial solution;  then
if we do the reverse operations on $D_c(1)$ with sensor $s_{r+2}$ by
using the gap list $G_1$, the solution obtained is
$D_c(l'+1,r+2)$.
\end{lemma}
\begin{myproof}
Consider the configuration $D_c(l',r+1)$. Since $D_c(1)$ is not a
trivial solution, we claim that the right
endpoint of $I(s_{r+1})$ in $D_c(l',r+1)$ must be at $\beta$. Indeed, if this is not
true, then according to the order preserving property, since $s_{r+1}$
is the rightmost sensor in $S(l',r+1)$,
the right endpoint of $I(s_{r+1})$ must be strictly to the right of
$\beta$, which implies that $I(s_{r+1})$ defines an overlap $o$ to the
right of $B$. Recall that our algorithm for computing $D_c(1)$ from
$D_c(l',r+1)$ is to cover $I(s_{l'})\cap B$ by eliminating overlaps of
$D_c(l',r+1)$ from left to right until $I(s_{l'})\cap B$ is fully
covered. Since $o$ is the rightmost overlap of $D_c(l',r+1)$ and
$S(l'+1,r+1)=\lambda>\frac{\beta}{2z}$, the overlap $o$ cannot be fully
eliminated in $D_c(1)$, which implies that $D_c(1)$ is a trivial solution,
incurring contradiction. Therefore, the above claim is proved.
	
According to our previous discussion, we can compute $D_c(l'+1,r+2)$
based on $D_c(0)$ in the following way. Suppose we have already
computed $D_c(0)$ and its gap list $G$. First, we compute $D_c(l',r+1)$
by doing reverse operations on $D_c(l')$ and $G$ with sensor
$s_{r+1}$, and $G_1$ is the list of remaining gaps of $G$.
Second, we compute $D_c(l',r+2)$ by
doing reverse operations on $D_c(l',r+1)$ and $G_1$ with sensor
$s_{r+1}$. Third, we remove $s_{l'}$ and cover $I(s_{l'})\cap B$ by
eliminating the overlaps of $D_c(l',r+2)$ from left to right until
$I(s_{l'})\cap B$ is fully covered. The obtained solution is
$D_c(l'+1,r+2)$. Note that the correctness of the first two steps is
based on our one-sided case algorithm, and that of the third step is
similar to Lemma \ref{lem:140}.
We use $A$ to denote the above algorithm for computing $D_c(l'+1,r+2)$.

Let $D$ be the configuration obtained after we do reverse operations
on $D_c(1)$ and sensor $s_{r+2}$ with the gap list $G_1$.
In summary, we obtain $D$ in the following way. Suppose we have already
computed $D_c(0)$ and its gap list $G$. First, we compute $D_c(l',r+1)$
by doing reverse operations on $D_c(l')$ and $G$ with sensor
$s_{r+1}$, and $G_1$ is the list of remaining gaps of $G$. Second, we
remove $s_{l'}$ and cover $I(s_{l'})\cap B$ by eliminating the gaps of
$D_c(l',r+1)$ from left to right until $I(s_{l'})\cap B$ is fully
covered. The obtained solution is $D_c(1)$. Third, we do reverse
operations on $D_c(1)$ and $s_{r+2}$ with $G_1$, and the obtained
solution is $D$. Let $A'$ denote our algorithm above.

Our goal is to prove that $D$ is $D_c(l'+1,r+2)$. To this end, we show
that each sensor of $S(l'+1,r+2)$ has the same location in $D$ and
$D_c(l'+1,r+2)$.

Both algorithms compute $D_c(l',r+1)$ after their first steps.
Let $o'$ be the rightmost overlap in $D_c(l',r+1)$. Recall that
we have proved that the right endpoint of $I(s_{r+1})$ in
$D_c(l',r+1)$ is at $\beta$. Hence, $o'$ cannot be an overlap to the
right of $\beta$.
Below, we assume $o'$ has two generators $g_k$ and $g_{k+1}$ since the
case where $o'$ has only one generator can be proved similarly but in
a simpler way. In the following discussion, in some configurations, the size of $o'$
may be changed but its generators are always $g_k$ and $g_{k+1}$;
for simplicity of discussion, we always use $o'$ to
refer to the overlap defined by $g_k$ and $g_{k+1}$ in any
configuration.

The second step of algorithm $A$ computes $D_c(l',r+2)$ by doing
reverse operations on $D_c(l',r+1)$ with $s_{r+2}$. As in the proof of
Lemma \ref{lem:100}, since $o'$ is an
overlap in $D_c(l',r+1)$, the result of the above reverse operations
only depends on the locations of the sensors of $S(k+1,r+1)$ in
$D_c(l',r+1)$, i.e., for each sensor $s_i\in S(k+1,r+2)$, its location
in $D_c(l',r+2)$ only depends on the locations of the sensors of
$S(k+1,r+1)$ in $D_c(l',r+1)$.

The second step of algorithm $A'$ computes $D_c(1)$ by removing
$s_{l'}$ and covering $I(s_{l'})\cap B$ by eliminating the
gaps of $D_c(l',r+1)$ from left to right.
We claim that the location of the sensor $s_{k+1}$ is the same in
$D_c(l',r+1)$ and $D_c(1)$. Indeed, since $|S(l',r+1)|=\lambda +1$,
$\lambda>\frac{\beta}{2z}$, and $2z\cdot |S(l',r+1)|>\beta+2z$, the
total length of the overlaps in $S(l',r+1)$ is strictly larger than
$2z$. Note that $|I(s_{l'})\cap B|\leq 2z$.
Since we cover $I(s_{l'})\cap B$ by eliminating the
overlaps of $D_c(l',r+1)$ from left to right (to obtain
$D_c(1)$) and $o'$ is the rightmost overlap of $D_c(l',r+1)$, $o'$
will not be fully eliminated in $D_c(1)$, which implies that $s_{k+1}$
will not be moved during the above procedure for covering
$I(s_{l'})\cap B$, i.e., $s_{k+1}$ has the same location in
$D_c(l',r+1)$ and $D_c(1)$. Further, due to the order preserving
property, each sensor of $S(k+1,r+1)$ has the same location in $D_c(l',r+1)$ and
$D_c(1)$.

With the above discussion, we prove below that each sensor
of $S(l'+1,r+2)$ has the same location in $D$ and $D_c(l'+1,r+2)$,
which will lead to the lemma.

\begin{enumerate}
\item
The second step of algorithm $A$ computes $D_c(l',r+2)$ by doing reverse
operations on $D_c(l',r+1)$ with $s_{r+2}$ and $G_1$; the third step
of algorithm $A'$
computes $D$ by doing reverse
operations on $D_c(1)$ with $s_{r+2}$ and $G_1$. We have discussed
above that the result of the reverse operations only depend on the
locations of the sensors of $S(k+1,r+1)$. Now that the locations of the sensors of
$S(k+1,r+1)$ are the same in $D_c(l',r+1)$ and $D_c(1)$, and
$o'$ exists in both configurations, the location of
each sensor of $S(k+1,r+2)$ must be the same in both $D_c(l',r+2)$ and
$D$.

\item
As discussed before, after the third step of algorithm $A'$
computes $D$ by doing reverse
operations on $D_c(1)$ with $s_{r+2}$, only sensors in $S(k+1,r+2)$
possibly change their locations. Therefore, each sensor of
$S(l'+1,k)$ has the same location in $D_c(1)$ and $D$.

\item
Since $o'$ exists in $D_c(1)$, $o'$ must exist in $D_c(l'+1,r+2)$. Indeed,
we can obtain $D_c(l'+1,r+2)$ by doing reverse operations on $D_c(1)$ and $s_{r+2}$. Hence, $o'$ must
exit in $D_c(l+1,r+2)$ although it may become longer (i.e., $s_{k+1}$ may be moved leftwards,
but $s_k$ does not change its location).

After the second step of algorithm $A$ computes $D_c(l',r+2)$ by the
reverse operations, $o'$ must exist in $D_c(l',r+2)$
although it may become longer that before.
Hence, each sensor of $S(l',k)$ has the same location in
$D_c(l',r+2)$ and $D_c(l',r+1)$.
The second step of algorithm $A'$ computes $D_c(1)$ by covering $I(s_{l'})\cap B$
by only moving the sensors in $S(l'+1,k)$ (because $o'$ still
exists in $D_c(1)$). The third step of algorithm $A$ computes $D_c(l'+1,r+2)$
by covering $I(s_{l'})\cap B$ by eliminating overlaps of $D_c(l',r+2)$
from left to right, in exactly the same way as $A'$ computes $D_c(1)$.
Since $o'$ exists in
$D_c(l'+1,r+2)$ and each sensor of $S(l',k)$ has the same location in
$D_c(l',r+2)$ and $D_c(l',r+1)$, algorithm $A$ can cover
$I(s_{l'})\cap B$ using the same sensors as does in $A'$.
This means
that each sensor of $S(l'+1,k)$ has the same location in
$D_c(1)$ and $D_c(l'+1,r+2)$.

\item
The third step of algorithm $A$ computes $D_c(l'+1,r+2)$
by covering $I(s_{l'})\cap B$ by eliminating overlaps of $D_c(l',r+2)$
from left to right. Since $o'$ exists in both $D_c(l',r+2)$ and $D_c(l'+1,r+2)$,
each sensor of $S(k+1,r+2)$ does not change its location in the above algorithm
for computing $D_c(l'+1,r+2)$, and in other words, each sensor of $S(k+1,r+2)$
has the same location in $D_c(l',r+2)$ and
$D_c(l'+1,r+2)$.
\end{enumerate}

To summarize our above discussion, we have obtained the following: (1)
each sensor of $S(k+1,r+2)$ has the same location in $D_c(l',r+2)$
and $D$; (2) each sensor of $S(l'+1,k)$ has the same location in $D_c(1)$ and $D$;
(3) each sensor of $S(l'+1,k)$ has the same location in
$D_c(1)$ and $D_c(l'+1,r+2)$; (4) each sensor of $S(k+1,r+2)$ has the
same location in $D_c(l',r+2)$ and $D_c(l'+1,r+2)$.

By the above (1) and (4), we obtain that each sensor of $S(k+1,r+2)$
has the same location in $D$ and $D_c(l'+1,r+2)$; by the above (2)
and (3), we obtain that each sensor of $S(l'+1,k)$
has the same location in $D$ and $D_c(l'+1,r+2)$. Therefore, each sensor
of $S(l'+1,r+2)$ has the same location in $D$ and $D_c(l'+1,r+2)$.
The lemma thus follows.
\end{myproof}

After we obtain $D_c(l'+1,r+2)$, we can use the same approach
to compute $D_c(2)$ (i.e., cover $I(s_{l'+1})\cap B$ by eliminating the overlaps
of $D_c(l'+1,r+2)$ from left to right).
We continue the same algorithm to compute $D_c(i)$
for $i=3,4,\ldots,l''$, until we find a trivial solution or $D_c(l'')$
is computed. We show in the following lemma that the entire
algorithm takes $O(n\log n)$ time.

\begin{lemma}\label{lem:170}
It takes $O(n\log n)$ time to compute $D_c(i)$ for $i=0,1,\ldots,l''$,
until we find a trivial solution or $D_c(l'')$ is computed.
\end{lemma}
\begin{myproof}
First, computing $D_c(0)$ can be done in $O(n\log n)$ time by our
containing case algorithm. We can also obtain the sets $G$, $O$, and
$O'$. Next, we use the algorithm discussed above to compute each
$D_c(i)$, which consists of two main steps.

On the one hand, recall that the first main step of computing each
$D_c(i)$ is to do reverse operations. Each reverse operation can be performed
in $O(\log n)$ time by updating the position tree $T_p$.
Recall that $q$ is the number of gaps in the gap list $G$ of $D_c(0)$.
The total number of the reverse operations in the entire algorithm
is at most $l''+q$, because after each reverse operation, either a gap is
removed from $G$ or a solution $D_c(l'+i,r+i+1)$ is obtained (as the
overlap defined by $s_{r+i+1}$ is eliminated during the operation).
Since $l''+q=O(n)$, the total time of the first main
steps in the entire algorithm is $O(n\log n)$.

On the other hand, the second main step of computing each $D_c(i)$ is
to cover $I(s_{l'+i-1})\cap B$ by eliminating the overlaps in the current list
of $O'$ from left to right. As discussed earlier, eliminating each
overlap takes $O(\log n)$ time by updating $T_p$. Hence, the
total time of the second main steps in the entire algorithm is
$O((l''+n_o)\log n)$, where $n_o$ is the
total number of overlaps that have ever appeared in
$O'$. Note that $n_o\leq n_o^1+n_o^2$, where $n_o^1$ is the number
of overlaps in $D_c(0)$ and $n_o^2$ is the number of overlaps restored
due to the reverse operations in the entire algorithm. Clearly, $n_o^1\leq
n$. Each reverse operation restores at most one overlap.
Hence, we have $n_o^2=O(n)$.
Thus, the total time of the second main steps in the entire algorithm
is $O(n\log n)$.

The lemma thus follows.
\end{myproof}

Recall that in the beginning
of this section we made an assumption that at least one sensor must
intersect $B$. In the case where the assumption does not hold,
we can use similar but much easier techniques to find an optimal solution in $O(n\log n)$ time, as shown in the lemma below.

\begin{lemma}\label{lem:180}
If the covering interval of every sensor of $S$ does not intersect $B$, then we can find an optimal solution in $O(n\log n)$ time.
\end{lemma}
\begin{myproof}
Suppose sensors in $S(1,k)$ are on the left side of $B$ and sensors in
$S(k+1,n)$ are on the right side of $B$. Hence, $x_k+z<0$ and $\beta+z<x_{k+1}$.

Since no covering interval intersects $B$ in the input configuration, due to the order preserving property, there must be an optimal solution $D_{opt}$ that uses a subset $S(l^*,r^*)$ of consecutive sensors to cover $B$ and the sensors of $S(l^*,r^*)$ are in attached positions. Further, sensors of $S\setminus S(l^*,r^*)$ are at their original locations.

Consider the configuration $D_{opt}$.
Since sensors of $S(l^*,r^*)$ are in attached positions and the covering interval of each sensor of $S(l^*,r^*)$ intersects $B$, the size of $|S(l^*,r^*)|$ is either $\lambda$ or $\lambda+1$. We claim that $|S(l^*,r^*)|$ cannot be $\lambda+1$. Indeed, assume to the contrary that $|S(l^*,r^*)|=\lambda+1$.
Clearly, either $|S(1,k)\cap S(l^*,r^*)|\leq |S(k+1,n)\cap S(l^*,r^*)|$
or $|S(1,k)\cap S(l^*,r^*)|> |S(k+1,n)\cap S(l^*,r^*)|$ holds.
Without loss of generality, we assume the former one holds.
Imagine that we shift all sensors of $S(l^*,r^*)$ rightwards.
Since $|S(l^*,r^*)|=\lambda+1$, during the above shift, at some moment the barrier
$B$ will be covered by the sensors in $S(l^*,r^*-1)$, i.e., sensor $s_{r^*}$ is redundant.
Further, due to $|S(1,k)\cap S(l^*,r^*)|\leq |S(k+1,n)\cap S(l^*,r^*)|$,
the above shift will not increase the value of $D_{opt}$.
Once $s_{r^*}$ becomes redundant, we stop the shift and move $s_{r^*}$ back to its original location in the input, which strictly decreases the value $D_{opt}$. This means that we obtain a solution that is strictly smaller than $D_{opt}$, contradicting with that $D_{opt}$ is an optimal solution.

Therefore, we obtain that $|S(l^*,r^*)|=\lambda$.
The above analysis can also show that there exists an optimal solution $D_{opt}$ with $|S(l^*,r^*)|=\lambda$ such that either the left endpoint of $I(s_{l^*})$ is at $0$ or the right endpoint of $I(l^*,r^*)$ is at $\beta$. Indeed, with loss of generality, we assume $|S(1,k)\cap S(l^*,r^*)|\leq |S(k+1,n)\cap S(l^*,r^*)|$ holds. If the left endpoint of $I(s_{l^*})$ is not at $0$ in $D_{opt}$, then we can always shift all sensors of $S(l^*,r^*)$ rightwards without increasing the value $D_{opt}$ until the left endpoint of $I(s_{l^*})$ is at $0$, at which moment we obtain an optimal solution in which the left endpoint of $I(s_{l^*})$ is at $0$.

Hence, there is an optimal solution with the following pattern:
(1) only sensors of $S(l^*,r^*)$ are moved and $|S(l^*,r^*)|=\lambda$;
(2) sensors  of $S(l^*,r^*)$ are in attached positions; (3) either $s_{l^*}$ is
at $z$ or $s_{r^*}$ is at $\beta-z$.

For any configuration $F$, here we define its {\em aggregate-distance} as the sum of the distances of all sensors between their locations in $F$ and their original locations in the input.

In light of the above discussion, to find an optimal solution,
we can do the following. First, we compute the aggregate-distances of the
 configurations for all $i=1,2\ldots,n-\lambda+1$ such that sensors of
 $S(i,i+\lambda-1)$ are in attached positions with $s_i$ at $z$.
 All these values can be computed in $O(n\log n)$ time by a ``sweeping'' algorithm
 similar to the one in Lemma \ref{lem:120}. Next, we compute the aggregate-distances
 of the configurations for all $i=1,2\ldots,n-\lambda+1$ such that sensors
 of $S(i,i+\lambda-1)$ are in attached positions with $s_{i+\lambda-1}$ at $\beta-z$. Similarly, this can be done in $O(n\log n)$ time. Finally, the configuration with the smallest aggregate-distance is an optimal solution to our problem.
\end{myproof}

The proof of the following theorem summarizes our algorithm for
solving the general case.

\begin{theorem}
The general case is solvable in $O(n\log n)$ time.
\end{theorem}
\begin{myproof}
We first check whether $|S_I|\geq \lambda$. If yes, by Lemma
\ref{lem:100}, it holds that $r^*=f(1)$. We can compute $f(1)$ by
applying our one-sided case algorithm on $S(1,n)$ after moving sensors
in $S_L$ rightwards to $-z$. After having $r^*$, as discussed earlier, we can find an
optimal solution in additional $O(n\log n)$ time, again by using our one-sided case algorithm.

Below we assume $|S_I|< \lambda$. If $|S_I|=\emptyset$, then we find an optimal solution by Lemma \ref{lem:180}.
Otherwise, we will compute two
candidate solutions $sol_1$ and $sol_2$, and the smaller one is our optimal solution.

Solution $sol_1$ corresponds to the case in Lemma \ref{lem:110}, i.e.,
$|S(l^*,r^*)|\geq \lambda+1$. By Lemma \ref{lem:110}, we have
$f(1)=\lambda^*$. Hence, we first compute $f(1)$ as above. Then, we
apply our one-sided case algorithm on the sensors of $S(1,f(1))$ after
sensors in $S(r+1,f(1))$ are moved leftwards to $\beta+z$, and the
obtained solution is $sol_1$.

Solution $sol_2$ corresponds to the case $|S(l^*,r^*)|=\lambda$.
If $\lambda=\frac{\beta}{2z}$, then we use the algorithm for Lemma
\ref{lem:120} to compute a solution of smallest value and the obtained
solution is $sol_2$. Otherwise, we compute all solutions $D(i)$ for
$i=0,1,\ldots,l''$ and return the smallest one as $sol_2$, which takes
$O(n\log n)$ time by Lemmas \ref{lem:150} and \ref{lem:170}.

Therefore, the total running time for computing $sol_1$ and $sol_2$ is $O(n\log
n)$.

The theorem thus follows.
\end{myproof}

\section{Concluding Remarks}
\label{sec:conclude}

In this paper, we present an algorithm that can solve the \msbc\
problem in $O(n\log n)$ time. To develop the algorithm, we discover many interesting
observations and propose new algorithmic techniques.
Since the \msbc\ problem is a
fundamental geometry problem, we suspect that our algorithm can find other
applications as well. Moreover, the observations we discovered and algorithmic
techniques we proposed in this paper may
be useful for solving other problems related to interval
coverage.

We can easily prove the $\Omega(n\log n)$ time lower bound for the \msbc\
problem (even for the containing case) by a reduction from the sorting
problem. Consider sorting a set of numbers $A=\{a_1,a_2,\ldots,a_n\}$.
In $O(n)$ time, we can create an instance for the \msbc\ problem as follows. Let
$S=\{s_1,s_2,\ldots,s_n\}$ be a set of $n$ sensors on the $x$-axis
$L$, and for each $1\leq i\leq n$, the coordinate of $s_i$
is $a_i$ and we say $s_i$ {\em corresponds to} $a_i$. Let $a'$
be the smallest number in $A$ and $a''$ be the largest number in $A$.
The barrier $B$ is the interval $[a',a'']$ on $L$. The covering range $z$ is
set to be $\frac{a''-a'}{2n}$. Clearly, this is an instance of the
containing case of the \msbc\ problem. Since $2z\cdot n$ is exactly
equal to the length of the barrier, the optimal solution has the
following pattern: all sensors are in attached positions and the
leftmost sensor is at $z$. Due to the order preserving property, the
left-to-right order of the sensors in the optimal solution corresponds
to the small-to-large order of the numbers in $A$. Therefore, once we
have the optimal solution, we can obtain the sorted list of $A$ in
additional $O(n)$ time. Since the sorting problem has $\Omega(n\log n)$ time lower
bound (in the algebraic decision tree model), the problem \msbc\ (even for
the containing case) also has $\Omega(n\log n)$ lower bound on the
time complexity.

\bibliographystyle{plain}
\bibliography{reference}

%




\end{document}